\documentstyle[aps,12pt]{revtex}
\def\Real{{\rm Re}}
\def\Imag{{\rm Im}}

\def\i{{\rm i}}
\def\e{{\rm e}}
\def\elect{e$^{-}$}
\def\h2o{H$_2$O}
\def\d2o{D$_2$O}

\def\mathcal{\cal}
\def\>{\rangle}
\def\<{\langle}

\begin{document}

\title{Numerical Simulations of Free Radical Dynamics}
\author{Eric R. Bittner\\
Department of Chemistry\\
University of Houston,\\ Houston TX, 77204 }
\date{\today}

\maketitle

\begin{quotation}
{\em Given for one instant an intelligence which could comporehend all the
forces by which nature is animated and the respective situation of the
beings who compose it--an intelligence sufficiently vast to submit
these data to analysis--it would embrace in the same formula the
movements of the greatest bodies of the universe and those of the
lightest atom; for it, nothing would be uncertain and the future, as
the past, would be present to its eyes.}--Laplace.
\end{quotation}

\section{Introduction}

Simulating chemical reactions in the condensed phase involves the 
parallel computation of the concerted quantum dynamics of breaking and 
formation of chemical bonds with the simultaneous response and 
influence of the surrounding media. In many systems it is not entirely 
possible to treat all of the motions quantum mechanically without 
resorting to various levels of approximation.  One approach which has 
proven to be quite powerful in a wide range of scenarios is to 
partition the full system into interacting subsystems
according to whether the dynamics in the subsystem is predominantly 
classical-like or if the dynamics must necessarily be treated 
quantum mechanically.   Such partitionings are motivated by the simple 
observation that quantum mechanical effects are typically localized 
to only a few specific types of motions typically along some chosen
reaction coordinate.  The remaining motions, i.e. those of 
the surrounding media, can largely be regarded as classical.  Thus, 
partitioning the extended system into interacting quantum and classical 
subsystems allows one to study the details of a condensed phase 
quantum phenomena while retaining a detailed description 
of the response and influence of a complicated molecular 
media in its full dimensionality.  
Consequently, the level of description afforded by mixed 
quantum/classical treatments and the relative computational efficiency 
of the method has allowed a synergistic interaction between theory and 
experiment since experiments can be use to validate theoretical models 
which in turn have provided a number of valuable mechanistic insights. 

Current quantum chemical methods have proven to be quite adequate in
providing structural and dynamical information for radicals in the gas
phase and inert matrices.  In these circumstances, quantum
computations have provided a direct link between experimental
spectroscopic investigations limited chiefly by the level of
sophistication needed to obtain accurate structures, spin dependent
properties such as hyperfine interactions, and higher order propertied
such as hyperpolarizabilities.

In the condensed phase, the story is quite different.  In order to
perform molecular dynamics of the solvent molecules, we need to
provide accurate empirical solvent-solvent and solute-solute
interaction potentials~\cite{MolDyn,AT}.  However, empirical models
cannot accurately reproduce many important modifications induced by
the solute on the electronic structure of the solvent molecules.
Consequently, the use of simple fixed point charge models for the
solvent ignores the polarization response of the solvent molecules to
changes in the electronic structure of the solute.  Mixed approaches
have been developed, in which a Monte Carlo simulation of the solvent
was coupled to a quantum mechanical description of the solute,
however, to date these have been performed at the Hartree-Fock level
due the computational overhead required to sample a statistically
meaningful number of solvent configurations\cite{HF1,HF2}.
Furthermore, the representation of the solvent molecules by simple
point charge models neglects many important polarization induced
effects when computing the electronic structure of the solute.

Solvent models have been developed in which the solvent is treated as
a polarizable dielectric continuum~\cite{PCM1,PCM2,PCM3,PCM4}.  While
the use of dielectric continuum models has a long history, the latest
generation of models are extremely sophisticated and include such
features as realistic solute cavities and exact description of the
electrostatic problem~\cite{Bartone96a,Bartone96b,Berne96}.  Such
Polarizable Continuum Models (PCM) have been included in the
Gaussian/94 (G94)~\cite{G94} package allowing an accurate extension of
a variety of quantum chemical approaches (including HF, MP2, CC, and
DFT) to condensed phase problems.  Such fully {\em ab initio}
approaches are currently accurate enough to interpret experimental
hyperfine splittings and permit an unbiased assessment of the role of
electronic, vibrational, and solvation effects in determining the
observed result~\cite{Bartone96b}.  The success of these models open
the possibility for studying a number of large biologically important
radical species in their natural aqueous environment.

{\em Ab initio} molecular dynamics methods, especially the Density
Functional Theory-Molecular Dynamics (DFT-MD) methods introduced by
Carr and Parrinello~\cite{MParrinello85,MParrinello89,MParrinello91}
(CP) have been highly successful in describing a variety of chemical
phenomena in the condensed
phase~\cite{MParrinello91,Parrinello93,Dhohl94}.  The CP method avoids
the necessity of providing semi-empirical pseudopotentials for the
molecular forces by solving the Kohn-Sham equations for the electronic
density for a given nuclear configuration.  The molecular dynamics are
then solved under the Born-Oppenheimer approximation in which the
nuclei move under the Hellmann-Feynman forces computed by the DFT
calculation.  Additional quantum effects in the nuclear motion, such
as energy level quantization and barrier tunneling, can be introduced
through path integral dynamics~\cite{MParrinello96a,MParrinello96b}.
While this methods is useful in determining mechanistic details of
many chemical reactions, electronic transition due to the
non-adiabatic motion of the nuclei cannot be properly include due to
the fact the Kohn-Sham equations are not guaranteed to be convergent
variationally for states other than the electronic ground state.
Thus, all nuclear dynamics are confined to the ground adiabatic
potential energy surface.  For a large number of systems of chemical
importance, this level of description is sufficient. However, for many
systems, sudden changes in the solute quantum state play pivotal roles
in determining the dynamics of the reaction.

Quantum-classical simulations overcome many fundamental shortcomings
associated with both classical molecular dynamics (MD) and the quantum
chemical methods mentioned above.  Foremost amongst the assumptions in
both MD and quantum chemistry is that all the dynamics occur on a
single potential energy surface under the Born-Oppenheimer
approximation.  Quantum mechanical effects arising from tunneling and
dynamical interference effects and discrete energy levels are
completely absent in classical MD.  These effects become significant
particularly in low temperature systems containing hydrogen atom
motions and for computing various time-dependent spectroscopic
observables.  The restriction of the dynamics to a single
Born-Oppenheimer potential surface is another serious limitation.  For
each non-degenerate electronic state of the many-atom system there is
a distinct Born-Oppenheimer potential energy surface parameterized by
the positions of all the atoms. When a transition between electronic
states occurs, the forces acting on the atoms may change dramatically.
Thus the proper and self consistent incorporation of these effects is
crucial to describing a wide host of chemical dynamical effects, many
of which are important to free radical dynamics.\cite{Yark96}

Electronically non-adiabatic processes play a surprisingly ubiquitous
role in chemical reaction dynamics for a wide number of systems
including photosynthesis, polymeric chain-reactions,
photodissociation, and electron transfer reactions.  and various
aspects which must be considered in order to simulate such processes
for a condensed matter system.  Foremost amongst the goals of this
chapter is to present the quantum-classical methodology as a  useful method
of simulating these processes in the condensed phase, to discuss some
of the subtle aspects of non-adiabatic dynamics which we have
demonstrated to be important considerations, and to present some of
our recent results obtained using these methods.

We consider here a system composed of a few important degrees of
freedom which include any reaction coordinate we care to choose
coupled to a much larger number of degrees of freedom which we shall
refer to as the bath.  We shall treat the motions of the system using
quantum mechanics and motions in the bath via classical mechanics.
The quantum motions may be, for example, nuclear motions, as in the
hydrogen transfer reaction, electronic degrees of freedom as in the
case of an electron transfer reaction or non-radiative relaxation
process, a combination of the two, and so forth.  One important class
of motions of chemical import which are inherently quantum mechanical
in nature are transitions between discrete electronic states. Since
the nuclear positions and velocities of all the molecules in the
system determine the electronic states and the transition amplitudes
between the states and the states and transition amplitudes in turn
influence the the forces governing the trajectories, any mixed quantum
classical theory must treat self-consistently the classical and
quantal degrees of freedom.

The aqueous electron, nature's simplest radical
anion,  provides an ideal test case system for the
methods which we shall develop below.  Many of the dynamical features
present in this system have analogues in other systems in which energy
quantization, non-radiative energy transfer, and other quantum effects
play import roles.  The aqueous electron is one of the very few
important chemical systems which can be described more or less exactly
without resorting to various levels of quantum chemical
approximations.  Thus, it provides a unique testcase for a variety of
dynamical theories.  

The remainder of this chapter is organized as
follows: 
we first present a general overview of classical molecular dynamics and
Monte-Carlo methods.  This is aimed as a general overview of the use
of molecular simulational methods.  As a part of this overview, we
discuss recent developments in {\em ab initio} molecular dynamics, in
particular the density functional theory 
methods developed by Carr and Parrinello.
We then focus our discussion on nonadiabatic simulations.
We 
present the formal details of quantum/classical dynamics
and highlight two important considerations which are necessary to
address: the treatment of interaction force and the treatment of the
phase coherence between the quantum and classical subsystems.  We then
move on to show results of simulations based upon these methods
focusing primarily upon the role that quantum decoherence plays in
determining the survival timescale of an excited electronic state
using the aqueous electron as a prototypical system,
highlighting the interplay which has evolved between theoretical and
experimental studies of this system over the past few years. Finally, we
conclude with a discussion of future directions we and others are
undertaking to apply the methods discussed here to a variety of other
systems.



\section{Excess Electron as a Prototypical Free Radical System}

The solvated electron provides perhaps the simplest example of
electronic dynamics in the condensed phase. Aqueous
electrons are of particular importance since they  play a prominent
role in the broad area of radiation chemistry of water.  In spite of
intensive experimental and theoretical study since identification of
the hydrated electron over 30 years ago, many dynamical features of
electron solvation have remained incompletely understood, largely due
to the extremely fast timescales on which electronic relaxation
occurs~\cite{Gaud87}.  Newly developed femtosecond spectroscopic
techniques are now providing glimpses of the details of condensed
phase electronic dynamics.  From a theoretical standpoint, the aqueous
electron is an ideal test case for new condensed phase theories.
Recent theoretical advances in treating condensed phase electronic
dynamics~\cite{PJR90,PJR91a,PJR91b,BJS94c,BJS94d} 
has led to an exceptional interplay between theory and experiment, helping to 
unravel many subtle dynamical aspects of this system.

While excess aqueous electrons are often prepared in conjunction with 
chemical reactions and charge transfer reactions in \h2o, 
experimental preparation of the electron in neat water can be 
accomplished by direct photoionization of water. Following 
injection into the water, the electron typically  has 2.0 to 2.5 eV 
of excess energy which must be absorbed by the water as the electron 
is solvated.  Initially, the electron is delocalized in a continuum 
band of energy levels.  However, the electron rapidly localizes 
into one of three nearly degenerate p-state
and subsequently relaxes to an s-state where it is further solvated 
by the surrounding water molecules.   This fully solvated species 
exists until scavenged by impurities in the system (typically on the 
order of a few hundred ns for ultra-pure water). The kinetics of the 
localization process has been extensively studied both theoretically 
and  experimentally. 

The fully solvated electron can be regarded as a particle in a 
hard roughly prolate spheroidal cavity formed by the 
surrounding water molecules.   The first ``solvation shell'' of 
water  molecules is an octahedral structure with hydrogen atoms 
on the vertices with an average e-/O radius of $\approx 3.2$ \AA.  
In the fully solvated state, 
the water molecules in the surrounding shells arrange 
themselves as to maximize the hydrogen bonding network. 
The first excited states of this system are
p-states lying roughly 1.7 eV above the ground state 
which are split due to slight asymmetry of the 
shell of water molecules about the electron.  The energy gap between 
the s and p states and the energies of the p states are modulated  
by fluctuations in the local  solvent structure. 

In the simulations which we shall discuss through the course of this 
chapter, we assume that the electron has been prepared in a 
equilibrated s-state and subsequently promoted into one of the p-states 
when the s-p energy gap becomes 1.7 eV simulating the 
effect of the 720 nm laser  excitation pulse used in the experiments 
performed by the Barbara and co-workers.
The relaxation mechanism following excitation is given by
\begin{eqnarray}
	e_{s} & \stackrel{h \nu}{\rightarrow} & e_{p}^{*}
	\stackrel{t_{1}}{\rightarrow} e_{p} \stackrel{t_{nr}}{\rightarrow}
	e_{s}^{*} \rightarrow e_{s}.
\end{eqnarray}
Following promotion into 
one of the p-states at $t=0$, the system is no longer in equilibrium and 
solvent water molecules move to accommodate the new distribution of 
electronic charge and partially solvates the p-state on a timescale, 
$t_{1}$.
During this time, the s-p energy gap to narrows from 1.7 eV to 
$\omega_{sp} \approx$ 0.5 eV. As the energy gap narrows, the 
non-radiative transition probability back to the ground state
increases and is maximized around avoided crossings.  
Thus, measurements of the excited state lifetime reflect both the
non-radiative decay of the state as well as the solvation/relaxation
dynamics of the surrounding media. 
The solvation dynamics of this system have been extensively studied 
by Schwartz and Rossky. The excited state solvation response, defined 
as the normalized autocorrelation function of the s-p energy gap 
following excitation,
\begin{eqnarray}
	S(t) & = & \left\langle \frac{\omega_{sp}(t) -
\omega_{sp}(0)}{\omega_{sp}(\infty)-\omega_{sp}(0)}\right\rangle,
\end{eqnarray}
indicates a rapid $\approx$ 24 fs inertial Gaussian component which 
makes up roughly 40\% of the response followed by a much slower 240 
fs  exponential decay which makes up the remainder. 
Following the non-radiative p $\rightarrow $ s 
transition, the new s-state ($e_{s}^{*}$) 
is no longer in equilibrium and the water 
solvent molecules respond to resolvate the electron.   
Figure \ref{fig:energy} shows the results of two typical 
runs following preparation in the excited state. 

An important puzzle concerning the non-adiabatic dynamics of
the hydrated electron has been the surprising lack of a sizable
isotope effect on the non-adiabatic transition rate.
\cite{Gaud91b,Barb94a} Estimating the 
non-radiative coupling between the excited and ground state 
using Fermi's Golden Rule 
\begin{eqnarray}
k_{10} \propto	|\vec{v}.\langle \psi_{0}|\vec{\nabla} \psi_{ex}\rangle|^{2}
\end{eqnarray}
shows
that the nuclear velocities, $\vec{v}$,
play a direct role in determination of the
non-adiabatic transition rate.  Since the fastest nuclear velocities
in D$_2$O (the D's) are classically $\sqrt{2}$ times slower than those in H$_2$O
(the H's)
while the other factors (the electron-water interaction potential, the
quantum force on the nuclei, etc.) remain the same,
the expectation is that non-adiabatic transition rates
should be roughly half as large in D$_2$O compared to H$_2$O. Indeed,
mixed quantum-classical simulations have suggested isotope effects of
factors of 2 $-$ 4 for the electronic transition rate in this
system.\cite{Nitzan91,Nitzan93,BJS96a} Experiments, however, have
found at most a modest difference in the non-adiabatic transition rate
for electrons photo-injected into H$_2$O versus
D$_2$O,\cite{Eisen90a,Eisen91,Gaud91a} and at most a 15\% 
isotopic effect
have been extracted from femtosecond spectroscopic studies of photoexcited
equilibrium electrons.\cite{Barb94a,Barb93a}
This disparity between theory and experiment comes as a surprise given the 
level of sophistication of the theories and their ability to match 
experiment on many other aspects of the system.  
This has caused us 
to re-evaluate the quantum-classical methodology and examine 
the various approximations used to derive the method,  which we shall 
describe in detail in the following section.




\section{Quantum-Classical Methodology}

\subsection{Non-adiabatic Dynamics}

Due to the fundamental role that dynamical process play in chemical
reactivity, there is a large body of literature devoted to extracting
dynamical information from computer simulations for both adiabatic and
non-adiabatic dynamics.
~\cite{PJR91a,PJR91b,ERB95b,Coker93,Tully71,Tully76,Tully90,PJR93,Tully94,Coker94,Coker95a}
As discussed above, it is not computationally feasible to treat a
condensed phase dynamical system fully quantum mechanically while
retaining a molecular level description of the dynamics; it is,
however, entirely possible to simulate the physics of interest by
treating a select few degrees of freedom quantum mechanically while
treating the remainder classically.  The general prescription goes as
follows: write the total Hamiltonian in terms of ``quantum'' and
``classical'' contributions,\footnote{Rigorously, one should start
from fully quantum realization of the system and introduce mixing
between the electronic and vibrational motions through the application
of the vibrational kinetic energy operator on the electronic
wavefunction to arrive at these equations. C. f. M. Born and K. Haung,
{\em Dynamical Theory of Crystal Lattices}, (Oxford Univ. Press,
1954).}
\begin{eqnarray}
H_{tot} &=& H_Q(x,R) + \frac{P^2}{2m} + V_C(R),
\end{eqnarray}
where $H_Q(x,R)$ is the Hamiltonian for just the quantum variables,
$x$, parameterized by the classical coordinates, $R$.  The other are
the classical kinetic energy and the solvent-solvent potential
interaction, $V_C(R)$, which is independent of the quantum subsystem.
The quantum subsystem is propagated in concert with the classical
particles using the time dependent Schr\"odinger equation, (note: we
set $\hbar = 1$ throughout)
\begin{eqnarray}
\i\left[\frac{\partial}{\partial t} + \sum_\mu\dot R_\mu
\frac{\partial}{\partial R_\mu} \right]\psi(x;R,t) & = & 
H_Q(x,R)\psi(x;R,t),
\label{eq:non_adi_se}
\end{eqnarray}
where the summation is over all classical coordinates.  Here we have
used the chain rule of differentiation to decompose the the total time
differential, to show explicitly the ``non-adiabatic coupling term'':
$\dot R \nabla_R \psi(x;R,t)$.  Under the Born-Oppenheimer
approximation, the quantum subsystem is assumed to respond much
faster than the timescale set by the motion of the classical
particles.  When this is the case, the non-adiabatic coupling term can
be ignored and the classical particles evolve on the adiabatic potential
energy surfaces defined by solving the time independent Schr\"odinger
equation at each classical configuration, $R$, for the adiabatic state
\begin{eqnarray}
H_Q(R,t) | \phi_{n}(R(t))\rangle =
 \varepsilon_n(R,t) | \phi_{n}(R(t))\rangle. 
\end{eqnarray}
Under the adiabatic approximation, the classical particles evolve 
under 
\begin{eqnarray}
m \ddot R_{\mu}(t) = -\frac{\partial }{\partial R_{\mu}} 
(\varepsilon_{n}(R(t)) + V_C(R(t))).
\end{eqnarray}
Thus, the coupling between solvent and solute are those predicted by 
the Hellmann-Feynman theorem
\begin{eqnarray}
F^{Q}_{\mu}(R(t)) &=& -\frac{\partial}{\partial R_{\mu}}
\langle \phi_{n}(R(t))| 
 H_{Q}(R(t))| \phi_{n}(R(t))\rangle 
\nonumber \\
&=& - \frac{\partial \varepsilon_{n}(R(t))}{\partial R_{\mu}}.
\end{eqnarray}

When the non-adiabatic coupling becomes significant, such as at an
avoided crossing of the adiabatic energy surfaces, the classical
motion can induce transitions in the quantum subsystem.  As the
classical particles move away from regions of strong non-adiabatic
coupling, the classical particles must evolve on a given adiabatic
surface.  This asymptotic condition introduces a twist on the
classical dynamics since the classical particles must ``switch''
surfaces during the course of a quantum transition.  Various
computational schemes have been developed which incorporate this
switching aspect into the classical dynamics.
Perhaps the most straightforward scheme is the ``surface hopping
algorithm'' pioneered by Tully~\cite{Tully71,Tully90} subsequently
incorporated in the work of Webster, Friesner, and
Rossky~\cite{PJR91a,PJR91b} and of Coker and
coworkers~\cite{Coker93,Coker94,Coker91,Coker92}. 
The general feature of all these algorithms
is that the classical particles stochastically ``switch'' between
quantum states according to a criteria determined by the quantum
transition probability between the initial and final quantum states
over a given time interval.  Thus, over the course of a simulation, a
switching trajectory emerges as the quantum system hops between the
various quantum states tracing out a path through state space. 

In mixed quantum-classical simulations, we assume that the nuclear
motion of at least a few degrees of freedom can be treated via
classical mechanics and that the evolution of remaining degrees of
freedom can be treated via quantum mechanics parameterized by
the classical variables.  The coherent evolution of the reduced
density matrix along a given classical trajectory, $\rho(R)$, is given
by the Liouville equation
\begin{eqnarray}
\i\frac{d\rho(R(t))}{d t}  &=& [ H_{Q}(R(t)),\rho(R(t))].
\end{eqnarray}
Here, the density matrix, $\rho(R(t))$, is defined along a single 
trajectory, $R(t)$ as
\begin{eqnarray}
	\rho(R(t)) & = & |\psi(R(t))\rangle\langle \psi(R(t))|,
\end{eqnarray}
where $|\psi\rangle$ is the solution of the time dependent 
Schr\"odinger equation. 
In the basis of adiabatic eigenstates 
this becomes
\begin{eqnarray}
    \rho(R(t)) &=& \sum_{ij}\rho_{ij}(R(t))
	|\phi_{i}(R(t))\rangle\langle \phi_{j}(R(t))|, 
\end{eqnarray}
with the time evolution of each component given by
\begin{eqnarray}
\i\dot\rho_{ij} &=& 
\left(\varepsilon_i\delta_{jk}-\i\dot R
d_{ik}\right)\rho_{kj}
-
\rho_{ik}
\left(\varepsilon_i\delta_{ik}-\i\dot R_{\mu}
 d_{kj}^{\mu}\right).
\end{eqnarray}
where $\varepsilon_i$ are the adiabatic energies 
computed for a given point
along the path, $d_{ij}^{\mu}$ is a component of the
non-adiabatic coupling vector
\begin{eqnarray}
d_{ij}^{\mu} = \langle \phi_i(R(t))|
\frac{\partial}{\partial R_{\mu}} |\phi_j(R(t))\rangle,
\end{eqnarray}
and  $\{|\phi_j(R(t))\rangle\}$ are the adiabatic eigenstates of
the quantum Hamiltonian at configuration $R(t)$.

Over a short time interval 
$\delta t$, we 
can construct the probability  for making a quantum 
mechanical transition from an initial 
state $|\phi_{i}(t)\rangle$ to a final state $|\phi_{f}(t+\delta 
t)\rangle$  writing
\begin{eqnarray}
P_{ij}(t_\circ+\delta t,t_\circ) &=& |\langle 
\phi_{j}|\exp(-\i\int_{t_{\circ}}^{t_{\circ}+\delta t}ds H(s))
\phi_{i}\rangle|^{2}\nonumber \\
&=&
\int_{t_\circ}^{t_\circ+\delta t}dt
\frac{\dot{\rho}_{jj}(t)}{\rho_{ii}(t)}\nonumber \\
 &\approx &  \delta t 
	\frac{\dot{\rho}_{jj}(t)}{\rho_{ii}(t)}
\end{eqnarray}
The first line is the definition of a quantum transition probability
The second and third lines are obtained by solving the equations of motion 
for the density matrix over a short time period which 
gives the probability of switching from
state $i$ to $j$ by the rate of change of the
population of the final state weighted by the population of the
initial state integrated over the short time interval. 
Using this we can write
\begin{eqnarray}
P_{ij}(t;\delta t)&=&
\delta t
\frac{-2 \Real \left\{\rho_{ji}(t)
\dot R(t)\langle \phi_i|
\frac{\partial}{\partial x}|\phi_j\rangle\right\}}{\rho_{ii}(t)}.
\end{eqnarray}
These probabilities are computed at each time step over the course of 
the simulation and determine whether or not the classical degrees of 
freedom will continue to evolve on the present energy surface or 
if they must switch energy surfaces due to a non-adiabatic transition in the 
quantum degrees of freedom.  
This particular approach for computing the 
quantum transition probability has the desired advantage that it 
minimizes the number of switches between states all the while 
maintaining the correct statistical distribution of state populations 
at all times.~\cite{Tully90} 

In the
Molecular Dynamics with Quantum Transitions (MDQT)
algorithm~\cite{Tully90,Tully94},
the classical variables switch suddenly between
the adiabatic potential surfaces at the switching times.
Such abrupt changes in the potential energy 
when an electronic transition does occur causes a change
in the total energy of the system. 
In the MDQT method, 
the velocity of the
classical variables are modified in order to properly conserve
energy~\cite{Tully90} by randomly 
rescaling the velocity 
vectors of the classical particles along the direction of 
non-adiabatic transition vector
such that the transition energy is transferred from the quantum to the 
classical degrees of freedom (or vise verse) in such a way that the total 
energy of the system is conserved.
We shall next discuss a theory in which the trajectories move 
on an effective potential which changes smoothly as the quantum state 
evolves from the initial to final state over a finite time interval.

\subsubsection{Solvent-Solute Coupling in the Presence of Quantum 
Transitions}

When a quantum transition does occur, the forces coupling the quantum 
system to the solvent change to reflect the new electronic state. 
This change is often dramatic, especially in transitions between an 
excited state and the ground electronic state. For example, in the 
case of an excess electron in \h2o, the first excited state is a 
p-state while the ground state is a roughly spherical s-state. 
Furthermore the partially solvated p-state occupies a much larger 
solvation cavity than the ground state for the same solvent 
configuration.  Thus, both the shape and spatial 
distribution of the electron changes suddenly as the result of an 
electronic transition.  Consequently, the forces coupling the electron 
to the solvent change suddenly as well.  A proper description of the
solvent-solute coupling must accommodate these sudden changes.

A further consideration is the fact that the total energy of the 
system must be conserved.  Any energy involved in making a transition 
in the quantum state must be absorbed or transferred to the solvent 
continuously throughout the transition.   In the case of an excess 
\elect in \h2o, the p-s transition dumps roughly 0.5 eV of electronic 
energy into the solvent.   This energy must be accommodated by the solvent 
translational, librational, and vibrational motions in a time scale 
of a few  femtoseconds. 


Along a given path, $R(t)$, the partial quantum 
transition amplitude between an initial and final quantum state,
$|i(R_i)\rangle$ and $|f(R_f)\rangle$ respectively, is given by
\begin{eqnarray}
T_{if}[R(t)] &=& \langle f (R_f) |
\e^{-\i\int_{t_i}^{t_f} ds H_Q[R(s)]}| \i (R_i) \rangle
\e^{+\i S[R(t)]}\nonumber \\ 
&=& U_{if}[R(t)]\e^{+\i S[R(t)]} ,~\label{tpart}
\end{eqnarray}
where $U_{if}[R(t)]$ is the transition amplitude for the
quantum subsystem and $S_{C}[R(t)]$ is the classical action
computed along the path $R(t)$, i.e.
\begin{eqnarray}
S_{C}[R(t)] = \int_{t_i}^{t_f} dt\left\{ \frac{m}{2}\dot{R}(t)^2 -
 V_C(R(t))\right\}.
\end{eqnarray}
We are calling this a {\em partial} amplitude here in the sense that 
this represents the full transition amplitude along one Feynman path 
taken by the solvent or bath degrees of freedom. 
Indeed, the full quantum transition probability of starting 
in some initial solvent/solute state
$|i(t_\circ)\rangle $ ending up in a given final solvent/solute
quantum state, $|f(t_f)\rangle $, 
at some time $t_f$ is computed by summing over all
possible Feynman paths, $R(t)$,
taken by the solvent between initial and 
final solvent configurations, $ R_i$ and $R_f$, and with initial and 
final electronic states, $|i(R_i)\rangle$ and $|f(R_f)\rangle$ respectively.
I.e.
\begin{eqnarray}
	K_{fi}(R_f, R_i; t) & = & \int {\cal D} R(t)
	 T_{if}[R(t)] \nonumber \\
	 &=& \int {\cal D} R(t)
	  \langle f (R_f) |
\e^{-\i\int_{t_i}^{t_f} ds H_Q[R(s)]}| \i (R_i) \rangle
\e^{+\i S_{C}[R(t)]}
\end{eqnarray}

The transition amplitude for the quantum subsystem, $U_{if}[R(t)]$, 
is a complex functional of $R(t)$.  We can obtain a semiclassical 
approximation for the combined system/bath propagator, $K_{fi}(R_f, R_i; t)$
on the assumption that the magnitude of $U[R]$ varies much more slowly 
than its phase along a given path. Expanding the phase to to second 
order in the classical paths $\tilde R$ yields the variational 
equation
\begin{eqnarray}
 \delta \left( \hbar \Imag \ln  U_{if}(\tilde R)
	 + S_{C}(\tilde R) \right) = 0.
\end{eqnarray}
The variation of the electronic propagator is given by
\begin{eqnarray}
	\delta U_{if}[R] & = & -\frac{i}{\hbar}\int_{t_{i}}^{t_{f}}ds 
	\langle f (t_{f}) | U(t_{f},s)
	\delta H_{Q}
	U(s,t_{i})
	|i(t_{i})\rangle.
\end{eqnarray}
Using this, we arrive at the ``classical'' equations of motion for 
the saddle point trajectories, $\tilde R$.
\begin{eqnarray}
	m \ddot{\tilde R}(t) & = & - \Real\left\{
	\frac{ \langle f (R_{f})| U(t_{f},t)\nabla H_{Q} 
	U(t,t_{i})|i(R_{i})\rangle}
	{\langle f (R_{f})| U(t_{f},t)|i(R_{i})\rangle}\right\}
	-\nabla V_{C}.
\end{eqnarray}
(We shall drop the tilde for now on.)
This is basically Newton's equations of motion.
Thus, the semiclassical estimate for the force contribution between
the quantum and classical degrees of freedom is given 
by~\cite{Pechukas69b}
\begin{eqnarray}
F_{Q}[R(t)]= -\Real\left\{
	\frac{ \langle f (R_{f})| U(t_{f},t)\nabla H_{Q} 
	U(t,t_{i})|i(R_{i})\rangle}
	{\langle f (R_{f})| U(t_{f},t)|i(R_{i})\rangle}\right\}. 
\end{eqnarray}
This force is a functional of the trajectory which we are trying to 
calculate over the course of a simulation and must be solved 
iteratively.  In other words, we guess a path and a set of end 
points, compute the quantum wavefunction for the new configuration,
propagate the initial wave forward in time and the final wave back in 
time to a set of intermediate times to compute the forces for a new 
path and continue with this cycle until either the procedure 
satisfies some convergence criteria or one simply runs out of 
patience or CPU time. 
This semiclassical/stationary phase 
description of the coupling between the quantum 
and classical variables is a central feature in the quantum/classical 
stationary phase surface hopping 
methods developed by Webster, Friesner, and Rossky (WFR)
\cite{PJR91a,PJR91b,PJR93}.
While providing an exact semi-classical description of the forces 
coupling the quantum and classical subsystems, the computational 
overhead required to iteratively determine the classical path can be 
quite large, especially if there are two or more paths with nearly 
identical action or if the time interval is taken to be too large. 
This creates considerable computational overhead since the
motion of all of the solvent  species must be determined 
variationally rather than through a propagative scheme, such as the 
Verlet algorithm.  

Furthermore, when we specify the initial and final 
states of the quantum solute, we are in effect imposing a quantum 
measurement on the solute which destroys coherence in the quantum 
subsystem. In the algorithm developed by Webster, Friesner, and 
Rossky (WFR), this ``computational coherence timescale'' is chosen to 
be identical to the timescale required to accurately compute the 
fastest motions in the solvent. Later developments by Coker 
and co-workers\cite{Coker93,Coker94,Coker91,Coker92}
and later by Bittner and Rossky~\cite{ERB95b,ERB96a,ERB96b,ERB97a}
have allowed for full retention and partial retention of coherence.

\subsubsection{Treating the phase coherence between the solvent and 
solute subsystems: Consistent Histories Formulation}

In both the MDQT and WFR/SP approaches, the classical switching
path can be written as a time ordered sequence of events,
\begin{eqnarray}
R^\alpha(t) = \{R_\circ^{\alpha_\circ},
\cdots, R_j^{\alpha_j} \cdots, R_n^{\alpha_n}\}\label{eq:path}
\end{eqnarray}
in which  superscript $\alpha_j$ denotes the switching outcome at time-step
$j$ and $\alpha_\circ = i$ corresponding to our choice of the initial
quantum state. Two such paths are shown schematically in
Fig.~\ref{fig:energy} where we plot the eigenenergy of the occupied state
along the path as a function of time for an aqueous electron following 
laser excitation.
Changes in the quantum state
imply that there is a corresponding sudden change in the forces
exerted on the classical particles over the course of its evolution.
The result is that different sequences of switching events will lead
to rapidly diverging paths.

Each path of the quantum subsystem will	have associated	with it	a
unique classical path characterizing the slight	differences	in the
bath dynamics arising from the slight differences in the forces	from
different quantum paths. These slight differences in the bath dynamics
mean that bath trajectories	will be	divergent even on a	very short
time scale.	 The result	of this	is that	the	off-diagonal elements of
the	quantum	density	matrix will	be rapidly diminished as the various
bath paths diverge.	 Thus, coherence in	the	quantum	subsystem is lost
due	to the slight differences in the bath dynamics for each	quantum
path.   We now explore the issue of phase	coherence between the
solvent	and	solute subsystems.
Our approach is based upon a novel interpretation of quantum mechanics
introduced by Griffiths~\cite{Griffiths84},
Omn{\`e}s~\cite{Omnes88,Omnes89}, and Gell-Mann and
Hartle~\cite{MGM90,MGM93,MGM94} over roughly the past 10 years.  This
interpretation of quantum mechanics, termed ``consistent'' or
``decoherent'' histories, has generated a great deal of attention in
the field of quantum
cosmology~\cite{Deco1,Deco2,Halliwell95a,Halliwell95b}.

Let us consider the evolution of an initial quantum state $| i
(R_\circ)\rangle$, taken as an adiabatic eigenstate of the quantum
mechanical Hamiltonian for initial bath (or classical) configuration
$R_\circ$, along a switching path, $R^\alpha(t)$.  During each time
interval, we determine whether or not a quantum transition has
occurred according to a stochastic switching criteria as discussed
above, modify the
classical dynamics accordingly, and record the
outcome of each switching attempt. 

We can write the reduced quantum transition probability for a mixed
quantum-classical system as
\begin{eqnarray}
P_{if}(t_f)& = & \left\langle \sum_{\{R^\alpha(t),R^{\beta}(t)\} }
T_{if}[R^\alpha(t)] T^\dagger_{if}[R^\beta(t)] \rho_i(R_\circ)
\right\rangle_{\circ}, \nonumber \\ &=& \left\langle
\sum_{\{R^\alpha(t),R^{\beta}(t)\} }
U_{if}[R^\alpha(t)]\rho_i(R_\circ))U^\dagger_{if}[R^\beta(t)] \e^{\i(
S[R^\alpha] - S[R^\beta])}\right\rangle_{\circ},
\label{eq:prob}
\end{eqnarray} 
where $\rho_i(R_\circ)$ is the probability density of being in the
initial state. The average, $\langle \cdots\rangle_\circ$,
is taken over initial configurations, and
the sum is over pairs of switching paths
\begin{eqnarray}
\{R^\alpha(t), R^\beta(t) | R^\alpha(0) = R_\circ, R^\beta(0) =
R_\circ\}
\end{eqnarray}
which start at the initial configuration, $R_\circ$, with the quantum
state in the initial state and end at any final configuration with the
quantum state in state $|f\rangle$ at $t_f$.

Notice that this definition of the transition probability no longer 
involves a single path.  In fact, the transition probability depends 
upon all other alternative paths which could ultimately lead to the 
final state.  
When the sequence of switching events for a pair of paths,
$R^\alpha(t)$ and $R^\beta(t)$ are identical, the contribution from
the classical action will vanish and the quantum transition
probability along a given path is given simply by
\begin{eqnarray}
P_{if}(t) =  \left\langle  
\sum_{\{R^\alpha(t)\}}| U_{if}[ R^\alpha(t)] |^2
\rho_i(R_\circ) \right\rangle_{\circ}.\label{eq:one_path}
\end{eqnarray}
which is identical to the transition probability we write in the 
previous section. 
Applying the assumption of Eq.~\ref{eq:one_path} to a switching path
implies that the coherences between the quantum states are not damped
by the bath. This can lead to the overestimation of the non-adiabatic
transition rates as demonstrated in Ref.~\cite{ERB95b}.

When transitions do occur, we must consider the phase interference
between the various alternative switching pathways.  At short times
following a switch, there will be a significant contribution from the
phase interferences between two alternative paths with similar
histories and the total transition probability is the sum over the
partial transition amplitudes for each pathway.~\cite{Feynman}
However, at longer times following a switch, the action difference
between two paths will be very large so that the 
phase interferences will be added  destructively,  leaving 
a sum over incoherent transition probabilities.  In short,
there will be considerable phase coherence between paths with similar
histories but little or no phase coherence between paths with
dissimilar histories.  The timescale over which we must explicitly
consider quantum interference effects between alternative pathways is
the {\em quantum decoherence time}.

Recently we demonstrated that quantum decoherence effects can be
consistently incorporated into mixed quantum-classical systems by
recognizing that restricting the quantum evolution to given classical
pathways is equivalent to making a series of quantum measurements on
the total system.~\cite{ERB95b,ERB96a,ERB96b} The classical path
sequence shown in Eq.~\ref{eq:path} and illustrated in Fig 1 is an
example of a quantum mechanical history. 
Along this history, the time evolution
operator for the quantum system can be written equivalently as a time
ordered sequence of alternating quantum projection operators and
unitary evolution operators
\begin{eqnarray}
\hat C[R^\alpha(t)]  =
\hat U_n \hat P^{\alpha_{n-1}} \cdots \hat U_1 P^{\alpha_1} \hat U_\circ
\end{eqnarray}
where the projections  at each time interval are members of complete
sets representing the total set of possible hopping outcomes,
\begin{eqnarray} 
\sum_{\alpha_i} \hat P^{\alpha_i} = 1,\label{pop1}
\end{eqnarray}
 and
\begin{eqnarray}
\hat U_n = \e^{-\i \int_{t_{n-1}}^{t_n}ds H[R^{\alpha}(s)]}
\end{eqnarray}
is the unitary evolution operator for the quantum wavefunction along a
segment of the switching path.  Every time we ``roll the dice'' to 
determine the next quantum state along the switching path, we need to 
apply the projection operator $\hat P$ to the quantum wavefunction.  

In general the projection operators in Eq.~\ref{pop1} are any 
operator acting on the Hilbert space of the quantum subsystem. In 
practice, we have found the Gaussian form to be useful.
\begin{eqnarray}
\hat P_i(Q) = \left({\frac{\alpha}{\pi}}\right)^{1/4}
e^{-\alpha(\hat Q - Q)^2/2} 
\end{eqnarray}
where $\hat Q$ is a bounded operator in the Hilbert space of the
quantum system with eigenvalues $\{Q_j\}$
Later we
shall define $\hat Q$ in terms of the coupling between the quantum and
classical variables.  The parameter $\alpha$ serves as the range
of values which are projected out by $\hat P$.
When acting on the quantum density matrix
$\hat P$ acts under the trace operation, 
\begin{eqnarray}
\hat P\rho  &=& \left(\frac{\alpha}{\pi}\right)^{1/2}
\int dQ \e^{-\alpha(\hat Q - Q)^2/2} \hat\rho
\e^{-\alpha(\hat Q - Q)^2/2} \nonumber \\
&=& \e^{-\alpha (Q_{i}-Q_{j})^{2}/4}\rho_{ij}.
\end{eqnarray}
Thus, when $\alpha\rightarrow\infty$, $\hat P$ projects out all
of the adiabatic eigenstate  and full coherence between the states is 
retained.  On the other hand, taking $\alpha = 0$ means that we 
project out only the state determined by the random switching event 
killing off any coherences between the eigenstates. 

In our approach, we use the decoherence timescale to set a
characteristic time interval between subsequent applications of the
projection operators.  During this interval, transition amplitudes are
added coherently according to the rules of quantum mechanics.
Application of the projection operators destroys the coherences and we
are left with transition probabilities which are then added according
to the rules of standard probability theory.  Within our approach,
Eq.~\ref{eq:prob} is rewritten as 
\begin{eqnarray}
P_{i\rightarrow f}(t) &=& \left\langle
\sum_{R^\alpha(t)} \langle f_n|
\hat C[R^\alpha(t)] \rho_i(R_\circ) 
\hat C^\dagger[R^\alpha(t)] | f_n\rangle
\right\rangle_{\circ} \nonumber \\
& =  & 
\left\langle 
\sum_{R^\alpha(t)}
\langle f_n | 
\hat U_n
\hat P^{\alpha_{n-1}}\cdots
\hat P^{\alpha_1}
\hat U_1
\rho_i^\circ
\hat U^\dagger_1
\hat P^{\alpha_1}\cdots
\hat P^{\alpha_{n-1}}
\hat U^\dagger_n
 | f_n \rangle
\right\rangle_{\circ},\label{eq:ch_prob}
\end{eqnarray} 
where $|f_n \rangle = |f(R^{\alpha_n})\rangle$ is the final quantum
state at the end of path $R^{\alpha}(t)$.

When the time intervals between ``measurements'',
$\delta t_n = t_n - t_{n-1}$, are members of
a Poisson distribution, the probability of maintaining
coherence over a given interval, $\delta t_{n}$, is 
\begin{eqnarray}
\pi(\delta t) = \int_0^{\delta t} dt \frac{\e^{-t/\tau}}{\tau}
= 1 - \e^{-\delta t/\tau}, \label{eq:poisson}
\end{eqnarray}
and $\tau$ is the characteristic decoherence timescale.  
Under these conditions,  the time evolution of the quantum
density matrix
over the short time step $\delta t$ is given as
\begin{eqnarray}
\rho(t+\delta t) = ( 1 - \frac{\delta t}{\tau})
(\rho(t)+\i\frac{1}{\tau} [H_{Q},\rho])
+ \frac{\delta t}{\tau} {\hat P}\rho.
\end{eqnarray}
Taking the limit of $\delta t \rightarrow 0$ yields the master equation 
for the quantum density matrix
\begin{eqnarray}
\dot{\rho} &=& i[H_{Q},\rho] 
 - \frac{1}{\tau} (\rho - {\hat P}\rho),\nonumber \\
&=&{\cal L}\rho - \frac{1}{\tau} (\rho - {\hat P}\rho).
\end{eqnarray}
Using the Gaussian projector above, 
\begin{eqnarray}
\dot{\rho}_{ij} &=& i[H_{Q},\rho] 
 - \frac{1}{\tau} (\rho_{ij} - [{\hat P}\rho]_{ij}),\nonumber \\
&=&[{\cal L}\rho]_{ij} - \frac{1}{\tau} (1-\exp[-\alpha 
(Q_{i}-Q_{j})^{2}/4])\rho_{ij}.
\label{eqn:lvn}
\end{eqnarray}
The first term contains the Liouvillian of the system, ${\cal L} =
i[H_{Q},\,]$. Evolution of the quantum system under this term alone is
unitary and non-dissipative.  The second term introduces {\em quantum
decoherence} into the dynamics of the quantum subsystem
due to the series of projections described above.  The
coherences originally in the quantum subsystem decay due to the
series of weak measurements imposed by the environment.

Both $\hat C[R^\alpha(t)] \rho_i(R_\circ) 
\hat C^\dagger[R^\alpha(t)] $ in Eq.~\ref{eq:ch_prob} and the 
density matrix in Eq.\ref{eqn:lvn} is a functional of only 
a {\em single} classical trajectory as opposed to a sum over
pairs of trajectories as in
Eq.~\ref{eq:prob}.  Where did the other paths go?  It turns out that
all pairs of paths are still included in these equations, except 
that we periodically ``prune'' the paths which branch away from the
main bundle of paths at either the switching times (when we determine 
a quantum transition to have occurred) or when we have determined 
that a ``measurement'' has occurred by randomly making the density 
matrix diagonal through application
of the projection operators
or by smoothly diminishing the off diagonal coherence using the 
master equation.  Both methods are formally equivalent.  

When the quantum system is coupled linearly in $\hat Q$
to a bath of oscillators, the fluctuations in the quantum system
become proportional to the ability of the bath to dissipate energy.
Under such conditions for a Markov bath, the relation
is~\cite{Zurek93}
\begin{eqnarray}
\frac{\alpha}{4\tau } = \frac{2 m\gamma_\circ kT}{\hbar^2},
\end{eqnarray}
where $\gamma_\circ$ is the dissipation constant.
One  can then relate decoherence in the quantum system to fluctuations
in the harmonic bath through the fluctuation-dissipation theorem. 
In
essence, our decoherence ansatz is entirely consistent with other
theories of quantum dissipation. In other theories of quantum
relaxation, such as the spin-boson model ~\cite{Leggett81,Leggett87},
master equations such as the Redfield relaxation
theory~\cite{Redfield,Abragam,Jean92,Jean94,Jean95}, or Liouville
space methods~\cite{MukamelNOS}, both decoherence and dissipation are
treated implicitly through effective equations of motion, thus losing
the molecular level information about the relaxation dynamics. In our
treatment, dissipation is included {\em explicitly} through the
classical molecular dynamics of the condensed phase medium, thus
retaining a molecular level description.  Quantum decoherence is
treated implicitly in order to avoid summing over alternative pairs of
classical paths.

\subsubsection{Decoherence Timescales}

Because our formalism relies upon an {\em a priori} knowledge of the
decoherence time scale for the given system, we desire  a
molecular level theory of the process, allowing access to a measure of
the quantum decoherence timescale. 
In developing a theory of quantum decoherence we introduced parameters
which characterize the decay of coherences due to differences in the
dynamics of the bath for each possible quantum state.  Here we shall
make use of the projection operator to solidify that concept and
determine how one can go about obtaining a measure of the coherence
time and length scales from realistic mixed quantum/classical
simulations.   

Let us consider the effect of coarse graining over a few
timesteps.  A the initial time $t_\circ$, we select the initial state
of the total system to be a pure state given by
$|\Psi_\circ(R_\circ)\rangle$.  Now, evolve $|\Psi_\circ\rangle$ to time
$t_1$ under the time dependent Hamiltonian $H(R(t))$ where $R(t)$ is a
classical path and resolve the final state in a basis
of eigenstates $\{|\phi^1_\alpha\rangle\}$ of $H(R(t_1))$.
\begin{eqnarray}
|\Psi_1(R_1)\rangle = \sum_{\alpha_1} 
c_{\alpha_1}|\phi^1_{\alpha_1}(R_1)\>.
\end{eqnarray}
Next, we project this state into coarse grained sets by writing
\begin{eqnarray}
\sum_{\alpha_1} c_{\alpha_1}|\phi^1_{\alpha_1}(R_1)\> =
\sum_{\Delta_1}\sum_{\alpha_1\in\Delta_1}
c_{\alpha_1}|\phi^1_{\alpha_1}(R_1)\>.
 \end{eqnarray} 
As a  result of the projection, there is little or no phase coherence 
between the various eigenstates composing this set.
Since each adiabatic
eigenstate defines a different adiabatic potential which the classical
path may follow from this point in time until the next time step, each
adiabatic state will define a unique coarse grained path or ``branch''
for the classical evolution.  Select a branch using some selection
criteria and propagate the state under the Hamiltonian,
$H(R_1(t))$, where $R_{1}(t)$ is a classical
trajectory along a given branch labeled by the sequence of adiabatic
potential surfaces followed by the trajectory. We accumulate these as an
indication of the history of choices made along the path.  Again at
time $t_2$, resolve the new state at $t_2$ in a basis of eigenstates
of $H(R_1(t_2))$ and select a new branch for the classical
dynamics
\begin{eqnarray}
\sum_{\alpha_1\in\Delta_1}
c_{\alpha_1}|\phi^1_{\alpha_1}(R_1)  \>
\mapsto 
\sum_{\Delta_2}\sum_{\alpha_2\in\Delta_2}
c_{\alpha_2}|\phi^{2\Delta_1}_{\alpha_2}(R_{21})\>.
\end{eqnarray}

Iterating the procedure once again to time $t_3$ yields the total
quantum state of the system/bath along a particular trajectory.
\begin{eqnarray}
\sum_{\alpha_2\in\Delta_2}
c_{\alpha_2}|\phi^{2\Delta_1}_{\alpha_2}(R_{21})  \>
\mapsto 
\sum_{\Delta_3}\sum_{\alpha_3\in\Delta_3}
c_{\alpha_3}|\phi^{3\Delta_2\Delta_1}_{\alpha_3}(R_{321})\>.
\end{eqnarray}

One can begin to see the conceptual underpinnings of the decoherent
histories approach to quantum mechanics.  At each level of coarse
graining, the evolution of the classical variables branches into
different possible histories where each branch indicates evolution
along a different adiabatic potential surface. The evolution of the
total quantum state depends very much upon the particular 
history, $\{\Delta_n,\Delta_{n-1},\cdots\}$, selected
at each time step.   In
other words, when we allow the classical variables to switch between
adiabatic potentials at a particular time as the result of a
electronic transition, the evolution of the electronic system becomes
dependent upon when the transitions actually occur.

Quantum coherence between alternative states is lost as the overlap
between the alternatives decays.   We can provide an estimate for the
rate at which coherence is lost considering the short time evolution
of the reduced density matrix along  a given path. For
example at time $t_3$, the reduced density matrix is given by
\begin{eqnarray}
\rho^{3\Delta_2\Delta_1}(R_{321}(t_3)) &=& 
\sum_{\Delta_3\Delta'_3}
\sum_{\alpha_3\in\Delta_3}\sum_{\beta_3\in\Delta'_3}
c_{\alpha_3}c_{\beta_3}^*\nonumber \\
&\times& |\phi_{\alpha_3}^{3\Delta_2\Delta1}
(R_{321})\rangle
\langle\phi_{\beta_3}^{3\Delta_2\Delta_1}(R_{321})|.\nonumber
\end{eqnarray}
To estimate the overlap between alternative branches for times $t >
t_3$ we assign to the bath coordinates at time $t_3$ Gaussian
wavepackets centered at the instantaneous phase space positions of the
bath variable, $|G(R,P)\rangle$.  In other words, we write
\begin{eqnarray}
|\phi_{\alpha_3}^{3\Delta_2\Delta1}
(R_{321})\rangle = |\phi_{\alpha_3}^{3\Delta_2\Delta1}\rangle
|G(R_{321},P_{321})\rangle.
\end{eqnarray}
Next, we make the approximation that we can
decouple the evolution of the quantum system from the
evolution of the bath over a short time period and allow the 
path to begin to branch in to $R_{321}$ and $R_{3'21}$.
\begin{eqnarray}
\rho^{3\Delta_2\Delta_1}(R_{321}(t>t_3)) &=& 
\sum_{\Delta_3\Delta'_3}
\sum_{\alpha_3\in\Delta_3}\sum_{\beta_3\in\Delta'_3}
c_{\alpha_3}(t)c_{\beta_3}^*(t)\nonumber \\
&\times&|\phi_{\alpha_3}^{3\Delta_2\Delta1}
(R_{321})\rangle 
\langle\phi_{\beta_3}^{3\Delta_2\Delta_1}(R_{3'21})|
\nonumber \\ &\times&
J_{\alpha_3\beta_3}(R_{321}-R_{3'21}; t).
\end{eqnarray}
$J_{\alpha_3\beta_3}(R_321-R_{3'21};
t)$ is the overlap integral between paths.
The decoherence time can thus be determined by the timescale at which
this vanishes. 

Let us approximate $J_{\alpha_3\beta_3}(t)$ using Gaussian coherent state
wavepackets which follow the classical evolution of the alternative
paths. Under this approximation, and to lowest order in time the
overlap decays  as (We drop the 321 and 3'21 notation and take $R= 
R_{321}$ at time $t_{3}$ and consider times $t\ge t_{3}$ ).
\begin{eqnarray}
J_{\alpha\beta}(R; t) &=&
\exp\left[-\frac{t^2}{4m\omega}(F^{\alpha}(R) -
F^{\beta}(R) )^2\right]
\end{eqnarray}
where $F^{\alpha}(R)$ and $F^{\beta}(R)$ are the instantaneous 
quantum forces
acting on the classical particles as the path bifurcates along
the various adiabatic potentials. (Note: that we are implicitly
summing over all particles in the system in the exponent of this 
equation.)  The width of the Gaussian,
$\omega$, is left arbitrary for the time being. 
Thus, a rough estimate of the quantum
decoherence time at a given configuration is given by the decay width of
$J_{\alpha\beta}(t)$,~\cite{ERB95b,ERB96a}
\begin{eqnarray}
\tau_D = 2\frac{\sqrt{m\omega}}{F^{\alpha} - F^{\beta}}.
\end{eqnarray}
Finally, averaging over an ensemble of configurations,
we obtain the ``average coherence decay'' 
\begin{eqnarray}
	\langle J(t)\rangle &=  & \int dR 
	P(R)\exp\left[-\frac{t^2}{4m\omega}(F^{\alpha}(R) -
F^{\beta}(R))^2\right]\label{javg},
\end{eqnarray}
where $P(R)$ is the probability distribution function for 
configuration $R$.

Such quantities  are easily computable in mixed
quantum/classical simulations and provide a convenient measure
for coherence times in simulations. We shall later use this
measure to assess the choice of decoherence times in simulations of
the dynamics of an excess electron in ordinary and heavy
water.~\cite{ERB96a}   The
coherence time for one bath configuration will be quite different than
the coherence time for another configuration.  
However, one can envision  a case in
which quantum transitions occur only when the classical variables are
in a few rare configurations.  In this case a global decoherence time
would not provide an accurate measure of the decoherence timescale.
As we shall demonstrate next, even small changes in the decoherence
timescale can have profound effects on both the quantum and the
classical evolution.

\subsubsection{Coherence Times For Physical Systems}

We can estimate
the decay of quantum coherence for the hydrated electron from
Eq. \ref{javg} with information available from excited state
simulations.  For the present example, the initial state $i$ is the
equilibrium excited state of the hydrated electron, and the final
state $j$ is the ground state of the electron.  For the widths of the
frozen Gaussians, we chose
\begin{eqnarray}
4 m \omega = 6 m k T \label{eqn22}
\end{eqnarray}
which results from rigorous analysis 
of the non-adiabatic transition rate
between displaced harmonic oscillators in the high temperature
limit,~\cite{Nitzan91,Nitzan93} and also allows for direct comparison to the
earlier calculations of Neria and Nitzan.~\cite{Nitzan91,Nitzan93} To compute
the average decoherence decay 
we constructed ensemble of excited state configurations 
by assuming that the system was equilibrated at
times past 1 ps after being prepared in the excited state. n
We chose 20 configurations at 25-50 fs intervals from
each of 5 individual trajectories for a total of 100 configurations.
Details of these simulations are given in the Appendix.

The results of this calculation are shown as in Fig.~\ref{fig:overlap}. 
For the hydrated electron the
coherence decays in a roughly Gaussian manner, and a Gaussian fit to
the decay has a variance of $\approx$ 3.1 fs.  Another estimate of the
decoherence time is found in the area under the curve, which for this
example is 2.8 fs.  This result is in good agreement with previous
calculations using a different model for the hydrated electron,
\cite{Nitzan91,Nitzan93} and 
demonstrates {\em a posteriori} justification for the hypothesis of a
$\approx$ 1 fs coherence time in the earlier non-equilibrium
simulations.  We also note that the rapid decoherence of this system
provides {\em a posteriori} justification for the short time
approximation used to estimate $\langle J(t)\rangle$.  We find that
the exponential is typically dominated by only 5 to 10 nuclei which
are the closest to the bulk of the electronic charge density.  This
makes sense from the stance that the largest difference in force
between the two surfaces will be for nuclei in positions where the
charge density, is large on one surface and small on the other.

We also show the coherence decay curve for an excited electron in 
\d2o computed similarly.
The coherence decay in D$_2$O is qualitatively similar to that in
H$_2$O, only for D$_2$O the approximate Gaussian decay time is
$\approx$ 4.6 fs (versus $\approx$3.1 fs for H$_2$O) and the area
under the curve is 4.1 fs (versus 2.8 fs for H$_2$O).

The longer coherence time in heavy water compared to light water
arises predominantly from the difference in mass in the choice of the
Gaussian width.  For classical H$_2$O and D$_2$O, the probability of a
given nuclear configuration is the same.  Static ensemble properties
for the two fluids should be identical since the ensembles contain
identical nuclear configurations with equal statistical weights.
\footnote{This is true for classical water models. When quantum 
mechanical effects are properly 
included, \h2o and \d2o have slightly different properties due to 
differences in the spatial dispersion of the H and D nuclei. 
C. f. G. S. Delbuono, P. J. Rossky, J. Schnitker, J. Chem. Phys. {\bf 
95}, 3728 (1991).} 
Since the electronic Hamiltonian for
the solvated electron is identical for both heavy and light water, the
static ensemble averaged potential energy difference and the
difference in Hellmann-Feynman forces on the two surfaces will also be
identical for the two fluids.  Thus, the only differences in the
evaluation of the coherence decay for the two fluids is 
through the mass term that enters through the Gaussian
width.  Since the nuclear overlap part of the
coherence decay depends on the sum over nuclei, the mass change leads
to the net slower decay of coherence in D$_2$O versus H$_2$O.  In
fact, for the purposes of evaluating $J(t)$ for the solvated electron
in D$_2$O, the H$_2$O simulations would suffice.

The different coherence decay times in the two solvents play a direct
role in determining the isotope effect on the overall non-adiabatic
transition rate.  In simplified terms, to determine the non-adiabatic
transition rate before quantum coherence has decayed, non-adiabatic
transition {\em amplitudes} should be added; after the decoherence
interval, memory of the complex phases is lost and non-adiabatic
transition {\em probabilities} should be added. 

\subsubsection{Effect of Decoherence on Quantum Transition Rates}
One of the
primary effects of electronic coherence loss can be observed in the
evaluation of the switching probabilities.  If we approximate the
short time evolution of the density matrix as
\begin{eqnarray}
\rho_{ij}(t+\delta t) &\approx&
\exp\left\{-\frac{ \delta t}{\tau} 
\left[
1 - \exp\left(-\alpha (F_i-F_j)^2/4\right)
\right]\right\}
\rho_{ij}(t),\label{eq:short-time}
\end{eqnarray}
and substitute this into the non-adiabatic switching probability
given above, one obtains
\begin{eqnarray}
\tilde P_{ij}(\delta t) &\approx&
\exp\left\{-\frac{\delta t}{\tau}
\left[
1 - \exp\left(-\alpha (f_i-f_j)^2/4\right)
\right]\right\}\nonumber \\
&\times&\int_{t}^{t + \delta t} ds \frac{-2 \Real \{ \rho_{ij} 
\dot R_{\mu} \langle \phi_{i}|\partial_{\mu}\phi_{j}\rangle\}}
{\rho{ii}}.
\end{eqnarray}
This predicts
that one should see a change over from non-adiabatic dynamics to adiabatic
dynamics as the coherence timescale becomes very small. 
Thus, even in this relatively simple model of coherence loss,
it is readily apparent that decoherence due to the coupling to an
external environment generally tends to suppress quantum effects and
increases the degree of adiabaticity in the quantum 
subsystem~\cite{ERB95b}.

As one can see from the last set of equations, one
can estimate the effect of coherence loss on the quantum transition 
probability given a coherence timescale and the
quantum transition amplitudes for a given system.  Recall that the 
origin of this equation is that the classical degrees of freedom (the 
bath) make a series of quantum measurements on the system at random 
time intervals.  Thus we can write the non-adiabatic transition 
probability between states $i$ and $j$ as a function of the 
measurement timescale $\tau$ as 
\begin{eqnarray}
	P_{ij}(\tau) & = & \frac{1}{Q}\sum_{N}\frac{1}{\tau}
	w_{N}(\tau, n \Delta t)\left|\sum_{n=1}^{\tau}T_{ij}^{(N)}(n \Delta 
	t)\right|^{2}.\label{ptau}
\end{eqnarray}
Here, $T_{ij}^{(N)}(n\Delta t)$ is the quantum transition amplitude at 
the $n^{\rm th}$ timestep starting from the $N^{\rm th}$ 
configuration.  In other words, to compute this quantity, we propagate 
the quantum subsystem coherently over a coherence time picked from a 
distribution of possible coherence times.  During this interval we sum 
over amplitudes. At the end of the interval,
coherences between states are killed off and we sum over 
probabilities.  The outer summation is over starting configurations 
and $w_{N}(\tau, n \Delta t)$ is the statistical weight for a given 
configuration to have a coherence time of $\tau$.  These are 
normalized by $Q$.   The time step, $\Delta t$ is taken as the 
shortest timestep used to propagate the equations of motion for 
the system.  For the case of the electron in \h2o and \d2o, this 
$\tau= 1$  fs throughout.

In Fig.~\ref{poisson}, we plot the estimated lifetime of the electron 
in the excited state as a function of coherence timescale for both 
solvents.  The solid lines, labeled  ``fixed'' refer to using a 
single coherence time interval in evaluating Eq.\ref{ptau}, i.e. 
$w_{N}(\tau, n \Delta t) /Q= \theta(\tau- n \Delta t)$. The curves 
labeled ``Poisson'' refer to lifetimes computed using a Poisson 
distribution of coherence time intervals. 
 \begin{eqnarray}
 	 w_{N}(\tau, n \Delta t)/Q& = & \exp(-n \Delta t/\tau)/\tau
 \end{eqnarray}
Both curves track each 
other quite well.  For a 1 fs coherence time, these lifetimes 
correspond to $\approx 550$ fs for \h2o and $\approx 850$ fs for 
\d2o.  The magnitude of these rates agree reasonably with the rates 
reported in earlier simulations~\cite{BJS94d,BJS94e,BJS95a} and the roughly 2:1 
simulated isotope effect is reproduced as well.  While non-adiabatic 
transitions typically occur from configurations with higher than 
average transition probabilities, these configurations are 
reasonably rare that the average transition probability determined 
here provides a reasonable estimate of the non-equilibrium population 
dynamics. 

We can now use the coherence timescales estimated above to provide an 
estimate of the excited state survival times.   For equal coherence 
times of 1 fs, which was the original assumption in the simulations by 
Schwartz and Rossky, the survival probabilities in Fig.\ref{poisson} 
predict roughly a 2:1 isotope effect.  However, when we use the 
decoherence times computed above, we find that the $\approx$ 50\% 
difference in coherence times yields an estimation of the lifetimes 
which are identical to within 10\%.  Furthermore, the lifetimes and 
estimated isotope effect agree remarkably with the experimental values 
reported by Barbara.  These results are summarized in Table 1. 
This provides a clear demonstration that quantum decoherence plays a 
direct and key role in the electronic dynamics of this system. 

\subsection{Decoherence in Quantum MD Simulations}

Quantum decoherence can be also be incorporated 
into non-adiabatic quantum MD simulations through 
either the projection operator method or the 
master equation theory discussed above.  In the statistical limit, 
both methods give identical results.~\cite{ERB95b} 
The computational 
algorithm given below uses the projection operator approach 
(Eq.\ref{eq:ch_prob}) with 
the coherence time intervals chosen from a Poisson distribution 
with characteristic time scale $\tau_{D} = 3.1 fs$ as derived 
from our estimates of the coherence time scale for an electron 
in \h2o when the excited state is nearly solvated. 
This time scale was estimated by computing the average 
decay time of the overlap of frozen Gaussian vibrational wavefunctions 
evolving on different adiabatic potential energy surfaces by sampling 
a large number of nearly equivalent excited state configurations.  
Also note that this method can be used for both the MDQT method and 
the WFR/SP methods since we assume that 
the treatment of the quantum-classical forces 
is independent of coherence retention.
The consistent histories algorithm proceeds as follows:

\begin{enumerate}
	\item Determine coherence interval from Poisson distribution 
	of possible intervals with characteristic time scale $\tau_{D}$.  
	\item Propagate the quantum wavefunction over this time scale while 
	self consistently evolving the classical degrees of freedom.  At 
	the end of each dynamical time step $h \le \tau_{D}$, determine the 
	switching path followed by the classical variables using the 
	stationary phase algorithm developed by Webster, {\em et 
	al.}~\cite{PJR91a,PJR91b} modified such that coherence in the quantum 
	wavefunction is maintained over the entire coherence interval.  We 
	also note that the stationary phase switching path is determined 
	in a piece-wise continuous fashion by selecting intermediate 
	quantum states every dynamical time step.  This is to avoid the 
	computational overhead of computing variationally the 
	stationary phase trajectory over relatively long time intervals.  
	So long as $\tau_{D}$ is no longer than a few dynamical time steps, 
	this approximation should not be too extreme; however, certain 
	pathological cases can be invented in which this approximation 
	does break down.
        
	\item If either a switch occurs in the time interval or we reach 
	the end of the interval, the quantum wavefunction is collapsed 
	using the projection operators discussed above.
	
	\item  Repeat.

\end{enumerate}

In Fig.~\ref{fig:swtimes} we plot the switching times from the excited 
state to the ground state for a total of 23 simulation runs.  The 
starting configurations were generated by performing a 35 ps 
simulation in which the electron was prepared in the ground state and 
16 initial configurations were chosen whenever the energy difference 
between the ground state and one of the p-like excited states become 
resonant with the excitation laser (1.7 eV). 
~\cite{Barb93a,Barb94a,Barbara93}
These starting 
configurations were identical to configurations used previously by 
Schwartz and Rossky in their work on this system.
~\cite{BJS94d,BJS94e,BJS95a}
 Following the 
initial excitation, the system was allowed to evolve.  During this 
time, the energy gap between the p-states and the s-state narrowed to 
$\approx 0.5 eV$ as the excited state was solvated by the surrounding 
water molecules.  The simulation continued until a switch from the 
excited state to the ground state was recorded.  Immediately after the 
switch, the energy gap between the occupied ground state and the first 
excited state widened dramatically as the solvent responded to the new 
electronic state.  The average and median switching times from these 
simulations are compiled in Table 2.

In order to test the sensitivity of the switching times to the choice 
of sequence of coherence intervals, 
five configurations were ``recycled'' by using different 
random number sequences for the coherence time intervals to generate 
different switching paths starting from the same initial configuration.
 For each such pairs of paths, the 
classical dynamics were identical up until the earlier switching time 
when one of the paths switched from the excited state to the ground 
state.  However, since the coherences between the different quantum 
states were killed off at randomly different times for each path, each 
path sampled a different probability distribution function for making 
a switch at each MD time step.  Hence any correlations between 
switching times resulting using the same initial configuration reflect 
correlations in the distribution of coherence time intervals.  Given 
that there is typically a 100 fs time difference between pairs of data 
(in one case a 530 fs difference) which is roughly 1/3 of the average 
survival time scale of the excited state, we find very little 
correlation between the switching times originating from the same 
initial configuration.

As the excited state is solvated by the surrounding water molecules, 
the energy gap between the excited state and the ground state narrows 
to its equilibrium value.  Furthermore, from first order perturbation 
theory, we expect that the electronic transition rate to be inversely 
proportional to the magnitude of the energy difference between initial 
and final states.  Thus, a simple model for the excited state survival 
probability can be written as
 \begin{eqnarray}
 	\frac{\partial P(t)}{\partial t} & = & -\frac{ k_{\rm eq}}{\tilde
\omega(t)}P(t)
 	\label{prob-eq}
 \end{eqnarray}
subject to the initial condition $P(0)=1$.  Here, $k_{\rm eq}$ is the 
non-adiabatic transition rate for the solvated excited state and 
$\tilde \omega(t)$ is the average energy gap normalized to the
average  energy gap of the solvated excited state.  $\tilde
\omega(t)$ is related  to the solvation response $S(t)$ by
\begin{eqnarray}
	\tilde \omega(t) & = &\frac{ \langle \omega_{\circ}\rangle 
	S(t) + \langle \omega_{\rm eq}\rangle (1-S(t))}{\langle \omega_{\rm
eq}\rangle}.
\end{eqnarray}
In Fig.~\ref{fig:gap}, we plot $\tilde \omega(t)$ using the fit
to the  solvation response function generated by our
simulations.  These fits are nearly identical to those
obtained by Schwartz and 
Rossky.\cite{BJS96a,BJS94c,BJS94d,BJS94e} At short  times,
when $\tilde \omega(t) > 1$, the non-adiabatic transition rate
will  be small since the energy gap is large.  At long
times $\tilde  \omega(t)\rightarrow 1$ as the energy gap relaxes
to its equilibrium value.  Thus, $k_{eq}$ is the
exponential decay constant of  the excited state population
once the excited state is fully solvated.

The solution of Eq.~\ref{prob-eq} is the one parameter family of curves
given by
\begin{eqnarray}
	P(t) & = & \exp\left(-k_{\rm eq}\int_{0}^{t} \frac{ds}{\tilde 
	E(s)} \right)P(0).
\end{eqnarray}
Using a non-linear fitting procedure, we fit our data to this model to 
obtain an estimate of the equilibrium non-adiabatic lifetime of 241 fs 
with a $\chi^{2}= 1.707 $.  A plot of our data superimposed on this 
fit is given in Fig.~\ref{fig:swtimes}.  Interestingly enough, our data 
does not fit this simple model as nicely as the data given by Schwartz 
and Rossky. This suggests that the
longer coherence times used in  this study imparts a non-trivial memory
dependency into the survival  probability which is inadequately captured
in Eq.~\ref{prob-eq}.

As expected, the lifetimes reported here (Table 2) 
are consistently shorter than 
the lifetimes reported by Schwartz and
Rossky
in which a constant 1 fs  coherence
time scale throughout their simulations, thus emphasizing the  profound
sensitivity of these simulations to the coherence time scale.   
Furthermore, our
results are consistent with the estimated lifetimes  estimated above.

\section{Discussion}

In this chapter we have briefly described the results of our work
towards a molecular level description of quantum relaxation phenomena.
Here we have focused exclusively upon the role that transient quantum
mechanical coherences between the solvent and the solute play in the
electronic relaxation of an excited solute species.  In mixed
quantum-classical computer simulations, fundamental assumptions about
the decay of these coherences produce direct manifestations on the
computed quantum mechanical transition rates and must be included
consistently in order to make realistic predictions and
comparisons.~\cite{ERB95b,ERB96a}

We have focused our attention primarily upon electronically
non-adiabatic processes and various aspects which must be considered
in order to simulate such processes for a condensed matter system.  In
most quantum-chemical calculations, the fundimental assumption is that
the nuclear dynamics are well within the Born-Oppenheimer
approximation and that break-downs in this approximation are rare and
unusual events.  However, in nature the opposite is largely true.
Non-adiabatic events occur in a number of important chemical processes
including the process of vision and photosynthesis, predissociation,
photodissociation, charge transfer, spin-quenching, and charge
transfer reactions of which free radical chemistry is a vital
component.  

The computational algorithm, based upon the so-called consistent
histories interpretation of quantum mechanics, provides both the
molecular level underpinnings of quantum decoherence and the
computational means for properly including decoherence effects in
non-adiabatic quantum-molecular dynamics simulations.  According to
the rules of ordinary quantum mechanics, a quantum system will evolve
into a coherent superposition of alternative states.  In our
decoherence theory, this coherence is dissipated due to the
differences in the forces exerted on the bath by alternative states
involved in the superposition.  Thus, on short scales, a quantum
system in a bath will obey the rules of ordinary quantum mechanics and
evolve into a coherent superposition of states, whereas on longer time
scales, the coherences between states are diminished and the quantum
system must be described as statistical (i.e., incoherent) mixture of
states.  As the quantum system interacts continuously with the bath,
coherences between states are continuously created by non-adiabatic
coupling and damped by the divergence in the bath dynamics induced by
the system-bath coupling.

This subtle interplay between coupling and decoherence and the
subsequent dependency of transition rates on the decoherence time
scale has profound implications for a variety of condensed phase
chemical dynamics including: internal conversion and internal
vibrational energy redistribution (in which the bath is comprised of
all the modes of the molecules except for the one mode of interest),
electronic energy transfer between molecules or different parts of the
same molecule, and charge transfer reactions including proton and
electron transfer.  In these latter examples, both the condensed phase
environment and the internal motions of the molecules act as a bath
which couples the quantum states together.  The decay of quantum
coherence, which depends upon the frequencies and populations of the
bath modes coupled to the quantum system will determine the extent to
which the non-adiabatic coupling can act to allow the chemical
reaction to proceed.  Changes in the spectral density due to isotopic
changes in the bath can have a substantial impact on non-adiabatic
chemical dynamics~\cite{ERB96a,ERB96b}.  Furthermore, the decay of
quantum coherence can determine the adiabaticity for a chemical
reaction.

Perhaps the two major lacunae in our theory of decoherence is the
explicit dependency upon an {\em a priori} estimate of the coherence
time scale and the fact that this time scale remains fixed throughout
the calculation.  As mentioned above, we estimated this time scale by
computing the average decay time of the overlap of a product of
Gaussian coherent states evolving on different adiabatic potential
energy surfaces.  In this estimate, the individual widths of the
coherent state wavefunctions centered about the initial phase space
points of the classical nuclei are set to be proportional to the
thermal DeBroglie wavelength of each nuclei.  Although physically
realistic for a variety of situations, this does leave the coherence
{\em length scale} (i.e.  the widths) as an adjustable parameter.
While the effects of changing the coherence length scale over a broad
range have not been systematically studied, results from our previous
work demonstrate that the non-adiabatic transition is quite sensitive
to changes in the coherence time scale and hence will be sensitive to
changes in the coherence length scales.  Furthermore, as the bath
explores various regions of the quantum potential energy surface, the
force differences which give rise to the decay of the quantum
coherences should vary from one configuration to the
next.~\cite{ERB96a} Current progress is underway towards obtaining
both the coherence length and time scale during the course of the
non-adiabatic simulation {\em ab initio} by examining the quantum
mechanical fluctuations of the bath particles about their stationary
phase paths.

\section*{acknowledgments}

This work was supported by the National Science Foundation as well as the San 
Diego Supercomputing Center for computing resources. 
We also thank Peter J. Rossky, Benjamin J. Schwartz, 
and Oleg V. Prezhdo for 
many fruitful discussions over the course of this work.   
The author also wishes 
to thank Hans Andersen for his hospitality at Stanford 
University where this chapter was completed.

\newpage

\begin{center}
{\bf Tables}
\end{center}

\begin{table}[h]
\caption{Summary of experimental and theoretical estimates of the
nonradiative lifetime of an excess electron in \h2o and \d2o.  The
first two rows are experimental results from Gauduel and Barbara and
the remainder are theoretical estimates. All theoretical numbers
reported are for relaxation from an {\em equilibrated} first excited
state of the electron in either solvent.  See text for details
regarding the differences in the theoretical values.}\label{table1}
\begin{tabular}{l|ccc}
\hline\hline
           &\h2o (fs)&\d2o(fs) &\d2o/\h2o \\
\hline
Gauduel {\em et al.} (Ref.~\cite{Gaud91b})  & 240 & 250  & 1.04 \\
Kimura {\em et al.} (Ref.~\cite{Barb94a})
                         & 310 $\pm$ 80 & 310$\pm$ 80 & 1.0  \\
Neria and Nitzan (Ref.~\cite{Nitzan93}) & 220 & 800  & $\approx$ 4 \\
Schwartz and Rossky (Ref.~\cite{BJS96a}) & 450 & 850  & $\approx$ 2 \\
1fs  Coherence$^a$       & 550 & 850  & 1.55  \\
Present Estimate$^b$     & 310 - 285 & 345 - 334 & 1.11 - 1.17  \\
\hline
\end{tabular}

$^a$ 1.0 fs coherence time for both \h2o and \d2o.

$^b$ Estimated lifetime ranges correspond to estimated ranges of the 
coherence time for both solvents. \h2o: 2.8 -3.1 fs. \d2o: 4.1-4.6 fs.
(c.f. Fig~\ref{fig:overlap}.)
\end{table}

\begin{table}[t]
\caption{Excited state lifetimes for the aqueous electron. (SR= 
	Schwartz and Rossky, J. Chem. Phys. {\bf 101} 6902 (1994).  
	SBPR = Schwartz, Bittner, 
	Prezhdo, and Rossky, J. Chem. Phys. {\bf 104} 4942 (1996)). 
The coherence times used 
	in each study are listed in parentheses. }
\protect\label{table2}
	
\begin{tabular}{c|lll}
	\hline
	  & Present (3.1 fs ) & SR (1 fs)  & SBPR (2.8-3.1 fs)$^{({\rm a})}$ \\
	\hline
	   Median    & 338 fs & 630 & ---  \\
	  Average    & 384    & 730 & ---   \\
	 Equilibrium & 234    & 450 & 310-270   \\ 
	\hline 
\end{tabular} 

(a.) Only Equilibrium lifetimes considered. 
\end{table}

\newpage

\begin{center}
{\bf Figure Captions}
\end{center}
\begin{figure}
\caption{Energy levels and occupations for two trajectories starting 
from identical water configurations.  The occupied state as a function 
of time is indicated by  $+$ and $\diamond$ for the two trajectories.}
\label{fig:energy}
\end{figure}

\begin{figure}
\caption{Overlap function averaged over 100 equilibrated excited state 
configurations for \h2o and \d2o.}
\label{fig:overlap}
\end{figure}

\begin{figure}
\caption{Effective Excited State Lifetimes vs. Quantum Decoherence
Time.  Shown for both \h2o and \d2o are estimates of the excited state
lifetimes using the estimates of the quantum coherence
timescales.  As discussed in the text, we estimate the coherence times
to be 2.8-3.1 fs for \h2o and 4.1-4.6 fs for \d2o. The corresponding
lifetimes are 310 - 285 fs and 345-334 fs for \h2o and \d2o
respectively.  The experimental range of 310$\pm$ 80 fs is from Ref.
[28]. See text for details.}\label{poisson}
\end{figure}

\begin{figure}
\caption{Distribution of switching times following initial excitation
for e$^{-}$/\h2o.  Diamonds ($\diamond$) represent the 
switching times from the simulations. 
Superimposed curves are the results of a Gaussian 
fit (B) and a kinetic  model (C).}\label{fig:swtimes}
\end{figure}

\begin{figure}
\caption{Plot of the normalized energy gap between the ground and 
occupied excited state following initial excitation.}\label{fig:gap}
\end{figure}


\begin{thebibliography}{10}

\bibitem{MolDyn}
D.~C. Rappaport,
\newblock {\em The Art of Molecular Dynamics Simulations} (Cambridge,
\newblock 1995).

\bibitem{AT}
M.~P. Allen and D.~J. Tildesley,
\newblock {\em Computer Simulations of Liquids} (Oxford University Press,
\newblock New York, 1987).

\bibitem{HF1}
H.~Takase and O.~Kikuchi,
\newblock J. Phys. Chem. {\bf 98}, 5160 (1994).

\bibitem{HF2}
H.~Takase and O.~Kikuchi,
\newblock Chem. Phys. {\bf 181}, 57 (1994).

\bibitem{PCM1}
J.~Tomasi and M.~Persico,
\newblock Chem. Rev. {\bf 94}, 2027 (1994).

\bibitem{PCM2}
J.~L. Rivail and D.~Rinaldi,
\newblock in {\em Computational Chemistry: Review of Current Trends}, edited by
  J.~Leszczynski, World Scientific, Singapore, 1995.

\bibitem{PCM3}
C.~J. Cramer and D.~J. Truhlar,
\newblock in {\em Continuum Solvation Models on Solvent Effects and Chemical
  Reactivity}, edited by O.~Tapia and J.~Bertran, Kluwar, Dordrecht, 1996.

\bibitem{PCM4}
S.~Miertus, E.~Scrocco, and J.~Tomasi,
\newblock Chem. Phys. Lett. {\bf 255}, 327 (1996).

\bibitem{Bartone96a}
V.~Bartone,
\newblock Chem. Phys. Lett. {\bf 262}, 201 (1996).

\bibitem{Bartone96b}
N.~Rega, M.~Cossi, and V.~Barone,
\newblock J. Chem. Phys. {\bf 105}, 11060 (1996).

\bibitem{Berne96}
J.~S. Bader and B.~J. Berne,
\newblock J. Chem. Phys. {\bf 104}, 1293 (1996).

\bibitem{G94}
M.~J. Frisch, G.~W. Trucks, H.~B. Schlegal, P.~M.~W. Gill, B.~G. Johnson, M.~A.
  Robb, J.~R. Cheeseman, T.~A. Keith, G.~A. Petersson, J.~A. Montgomery,
  J.~Cioslowski, B.~B. Stefanov, A.~Nanayakkara, M.~Challacombe, C.~Y. Peng,
  P.~Y. Aylal, W.~Chen, M.~W. Wong, J.~L. Andres, E.~S. Replogle, R.~Gomperts,
  R.~L. Martin, D.~J. Fox, J.~S. Binkley, D.~J. DeFrees, J.~P. Stewart,
  M.~Head-Gordon, C.~Gonzalez, and J.~A. Pople,
\newblock {\em GAUSSIAN 94 (Revision B.3)} (Gaussian Inc., Pittsburgh,
\newblock 1994).

\bibitem{MParrinello85}
R.~Car and M.~Parrinello,
\newblock Phys. Rev. Lett. {\bf 55}, 2471 (1985).

\bibitem{MParrinello89}
R.~Car and M.~Parrinello,
\newblock in {\em Simple Molecular Systems at Very High Density}, edited by
  A.~Poliani, P.~Loubeyre, and N.~Boccra, Plenum, New York, 1981.

\bibitem{MParrinello91}
G.~Galli and M.~Parrinello,
\newblock in {\em Computer Simulations in Material Science}, edited by M.~Meyer
  and V.~Pontikis, Kluwer, Dordrecht, 1991.

\bibitem{Parrinello93}
K.~Laasonen, M.~Sprik, and M.~Parrinello,
\newblock J. Chem. Phys. {\bf 99}, 9080 (1993).

\bibitem{Dhohl94}
D.~Hohl and R.~O. Jones,
\newblock Phys. Rev. B {\bf 50}, 17047 (1994).

\bibitem{MParrinello96a}
D.~Marx and M.~Parrinello,
\newblock J. Chem. Phys. {\bf 104}, 4077 (1996).

\bibitem{MParrinello96b}
M.~E. Tuckerman, D.~Marx, M.~L. Klein, and M.~Parrinello,
\newblock J. Chem. Phys {\bf 104}, 5579 (1996).

\bibitem{Yark96}
D.~R. Yarkony,
\newblock J. Phys. Chem. {\bf 100}, 18612 (1996).

\bibitem{Gaud87}
A.~Migus, Y.~Gauduel, J.~L. Martin, and A.~Antonetti,
\newblock Phys. Rev. Lett. {\bf 58}, 1559 (1987).

\bibitem{PJR90}
P.~J. Rossky,
\newblock J. Opt. Soc. Am. B {\bf 7}, 1727 (1990).

\bibitem{PJR91a}
F.~Webster, P.~J. Rossky, and R.~A. Friesner,
\newblock Comput. Phys. Commun. {\bf 63}, 494 (1991).

\bibitem{PJR91b}
F.~Webster, J.~Schnitker, M.~S. Friedrichs, R.~Friesner, and P.~Rossky,
\newblock Phys. Rev. Lett. {\bf 66}, 3172 (1991).

\bibitem{BJS94c}
P.~J. Rossky, B.~J. Schwartz, and W.~S. Sheu,
\newblock Electronic relaxation dynamics in solution,
\newblock in {\em Ultrafast Phenomena IX}, edited by P.~F. Barbara, W.~H. Knox,
  G.~A. Mourou, and A.~H. Zewail, volume~60 of {\em Springer Series in Chemical
  Physics}, page~53 (Springer-Verlag, Berlin, 1994).

\bibitem{BJS94d}
B.~J. Schwartz and P.~J. Rossky,
\newblock J. Chem. Phys. {\bf 101}, 6902 (1994).

\bibitem{Gaud91b}
Y.~Gauduel, S.~Pommeret, A.~Migus, and A.~Antonetti,
\newblock J. Phys. Chem. {\bf 95}, 533 (1991).

\bibitem{Barb94a}
Y.~Kimura, J.~C. Alfano, P.~K. Walhout, and P.~F. Barbara,
\newblock J. Phys. Chem. {\bf 98}, 3450 (1994).

\bibitem{Nitzan91}
E.~Neria, A.~Nitzan, R.~N. Barnett, and U.~Landman,
\newblock Phys. Rev. Lett. {\bf 67}, 1011 (1991).

\bibitem{Nitzan93}
E.~Neria and A.~Nitzan,
\newblock J. Chem. Phys. {\bf 99}, 1109 (1994).

\bibitem{BJS96a}
B.~J. Schwartz and P.~J. Rossky,
\newblock J. Chem. Phys. {\bf 105}, 6997 (1996).

\bibitem{Eisen90a}
F.~H. Long, H.~Lu, and K.~B. Eisenthal,
\newblock Phys. Rev. Lett. {\bf 64}, 1469 (1990).

\bibitem{Eisen91}
F.~H. Long, H.~Lu, X.~Shi, and K.~B. Eisenthal,
\newblock Chem. Phys. Lett. {\bf 185}, 47 (1991).

\bibitem{Gaud91a}
S.~Pommeret, A.~Antonetti, and Y.~Gauduel,
\newblock J. Amer. Chem. Soc. {\bf 113}, 9105 (1991).

\bibitem{Barb93a}
J.~C. Alfano, P.~K. Walhout, Y.~Kimura, and P.~F. Barbara,
\newblock J. Chem. Phys. {\bf 98}, 5996 (1993).

\bibitem{ERB95b}
E.~R. Bittner and P.~J. Rossky,
\newblock J. Chem. Phys. {\bf 103}, 8130 (1995).

\bibitem{Coker93}
D.~Coker,
\newblock Computer simulation of non-adiabatic dynamics in condensed systems,
\newblock in {\em Computer Simulation in Chemical Physics}, edited by M.~Allen
  and D.~Tildesley, pages 315--377, Kluwer Academic Publishers, Dordrecht,
  1993.

\bibitem{Tully71}
J.~Tully and R.~Preston,
\newblock J. Chem. Phys. {\bf 55}, 562 (1971).

\bibitem{Tully76}
J.~Tully,
\newblock Non-adiabatic processes in molecular collisions,
\newblock in {\em Dynamics on Molecular Collisions, Part B}, edited by
  W.~Miller, page 217, Plenum, New York, 1976.

\bibitem{Tully90}
J.~C. Tully,
\newblock J. Chem. Phys. {\bf 93}, 1061 (1990).

\bibitem{PJR93}
T.~H. Murphrey and P.~J. Rossky,
\newblock J. Chem. Phys. {\bf 99}, 515 (1993).

\bibitem{Tully94}
S.~Hammes-Schiffer and J.~C. Tully,
\newblock J. Chem. Phys. {\bf 101}, 4657 (1994).

\bibitem{Coker94}
L.~Xiao and D.~Coker,
\newblock J. Chem. Phys. {\bf 100}, 8646 (1994).

\bibitem{Coker95a}
L.~Xiao and D.~F. Coker,
\newblock J. Chem. Phys. {\bf 102}, 496 (1995).

\bibitem{Coker91}
B.~Space and D.~Coker,
\newblock J. Chem. Phys. {\bf 94}, 1976 (1991).

\bibitem{Coker92}
B.~Space and D.~Coker,
\newblock J. Chem. Phys. {\bf 96}, 652 (1992).

\bibitem{Pechukas69b}
P.~Pechukas,
\newblock Phys. Rev. {\bf 181}, 174 (1969).

\bibitem{ERB96a}
B.~J. Schwartz, E.~R. Bittner, O.~V. Prezhdo, and P.~J. Rossky,
\newblock J. Chem. Phys. {\bf 104}, 5942 (1994).

\bibitem{ERB96b}
E.~R. Bittner, B.~J. Schwartz, and P.~J. Rossky,
\newblock J. Mol. Struct.:THEOCHEM  (1996),
\newblock in press.

\bibitem{ERB97a}
E.~R. Bittner and P.~J. Rossky,
\newblock J. Chem. Phys.  (1997),
\newblock in press.

\bibitem{Griffiths84}
R.~B. Griffiths,
\newblock J. Stat. Phys. {\bf 36}, 219 (1984).

\bibitem{Omnes88}
R.~Omn{\'e}s,
\newblock J. Stat. Phys. {\bf 53}, 893 (1988).

\bibitem{Omnes89}
R.~Omn{\'e}s,
\newblock Ann. Phys. {\bf 201}, 354 (1989).

\bibitem{MGM90}
M.~Gell-Mann and J.~B. Hartle,
\newblock Quantum mechanics in the light of quantum cosmology,
\newblock in {\em Complexity, Entropy and the Physics of Information}, edited
  by W.~H. Zurek, Addison-Wesley, Redwood City, CA, 1990.

\bibitem{MGM93}
M.~Gell-Mann and J.~B. Hartle,
\newblock Phys. Rev. D {\bf 47}, 3345 (1993).

\bibitem{MGM94}
M.~Gell-Mann and J.~Hartle,
\newblock Equivalent sets of histories and multiple quasi-classical domains,
\newblock LANL preprint: gr-qc:940413, 1994.

\bibitem{Deco1}
S.~Schreckenberg,
\newblock J. Math. Phys. {\bf 36}, 4735 (1996).

\bibitem{Deco2}
F.~Dowker and A.~Kent,
\newblock On the consistent histories approach to quantum mechanics,
\newblock LANL preprint: gr-qc/9412067, 1994.

\bibitem{Halliwell95a}
L.~Do{\'o}si, N.~Gisin, J.~Halliwell, and I.~C. Percival,
\newblock Phys. Rev. Lett. {\bf 74}, 203 (1995).

\bibitem{Halliwell95b}
J.~Halliwell and A.~Zoupas,
\newblock Phys. Rev. D  (1995),
\newblock (submitted).

\bibitem{Feynman}
R.~P. Feynman and A.~R. Hibbs,
\newblock {\em Quantum mechanics and path integrals} (McGraw-Hill,
\newblock New York, 1965).

\bibitem{Zurek93}
W.~H. Zurek, S.~Habib, and J.~P. Paz,
\newblock Phys. Rev. Lett. {\bf 70}, 1187 (1993).

\bibitem{Leggett81}
A.~O. Calderia and A.~J. Leggett,
\newblock Phys. Rev. Lett. {\bf 46}, 211 (1981).

\bibitem{Leggett87}
A.~J. Leggett, S.~Chakravarty, A.~T. Dorsey, M.~P.~A. Fisher, A.~Garg, and
  W.~Zwerger,
\newblock Rev. Mod. Phys. {\bf 59}, 1 (1987).

\bibitem{Redfield}
A.~G. Redfield,
\newblock Adv. Mag. Reson. {\bf 1}, 1 (1965).

\bibitem{Abragam}
A.~Abragam,
\newblock {\em The Principles of Nuclear Magnetism} (Oxford University Press,
\newblock London, 1961).

\bibitem{Jean92}
J.~M. Jean, R.~A. Friesner, and G.~R. Fleming,
\newblock J. Chem. Phys. {\bf 96}, 5827 (1992).

\bibitem{Jean94}
J.~M. Jean,
\newblock J. Chem. Phys. {\bf 101}, 10464 (1994).

\bibitem{Jean95}
J.~M. Jean,
\newblock J. Chem. Phys. {\bf 103}, 2092 (1995).

\bibitem{MukamelNOS}
S.~Mukamel,
\newblock {\em Principles of Nonlinear Optical Spectroscopy} (Oxford University
  Press,
\newblock New York, 1995).

\bibitem{BJS94e}
B.~J. Schwartz and P.~J. Rossky,
\newblock J. Chem. Phys. {\bf 101}, 6917 (1994).

\bibitem{BJS95a}
B.~J. Schwartz and P.~J. Rossky,
\newblock J. Mol. Liquids {\bf 65/66}, 23 (1995).

\bibitem{Barbara93}
J.~C. Alfano, P.~K. Walhout, Y.~Kimura, and P.~F. Barbara,
\newblock J. Chem. Phys. {\bf 98}, 5996 (1993).

\bibitem{Rahman85}
K.~Toukan and A.~Rahman,
\newblock Phys. Rev. B {\bf 31}, 2643 (1985).

\bibitem{PJR87}
P.~J. Rossky and J.~Schnitker,
\newblock J. Chem. Phys. {\bf 86}, 3462 (1987).

\end{thebibliography}
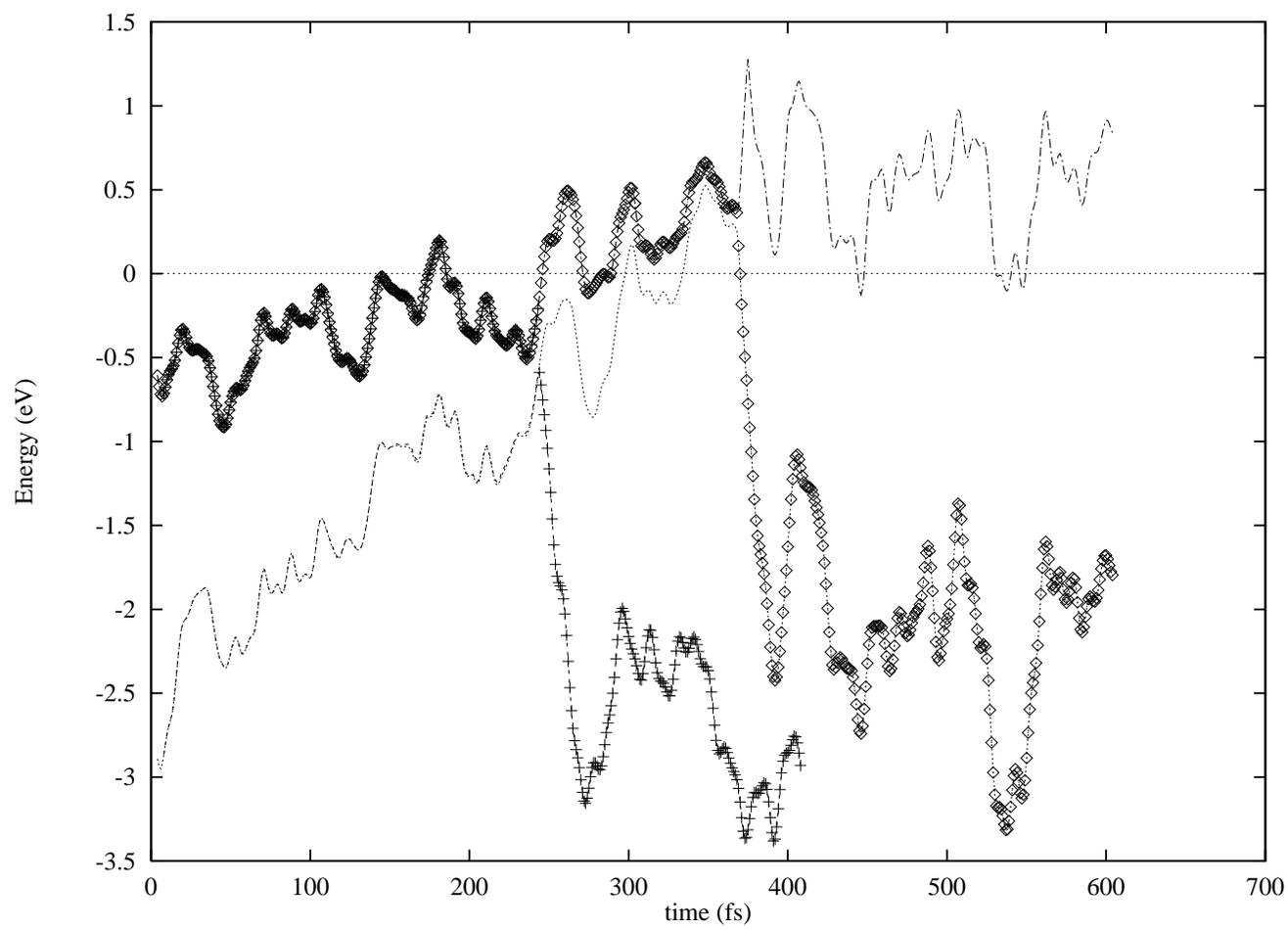
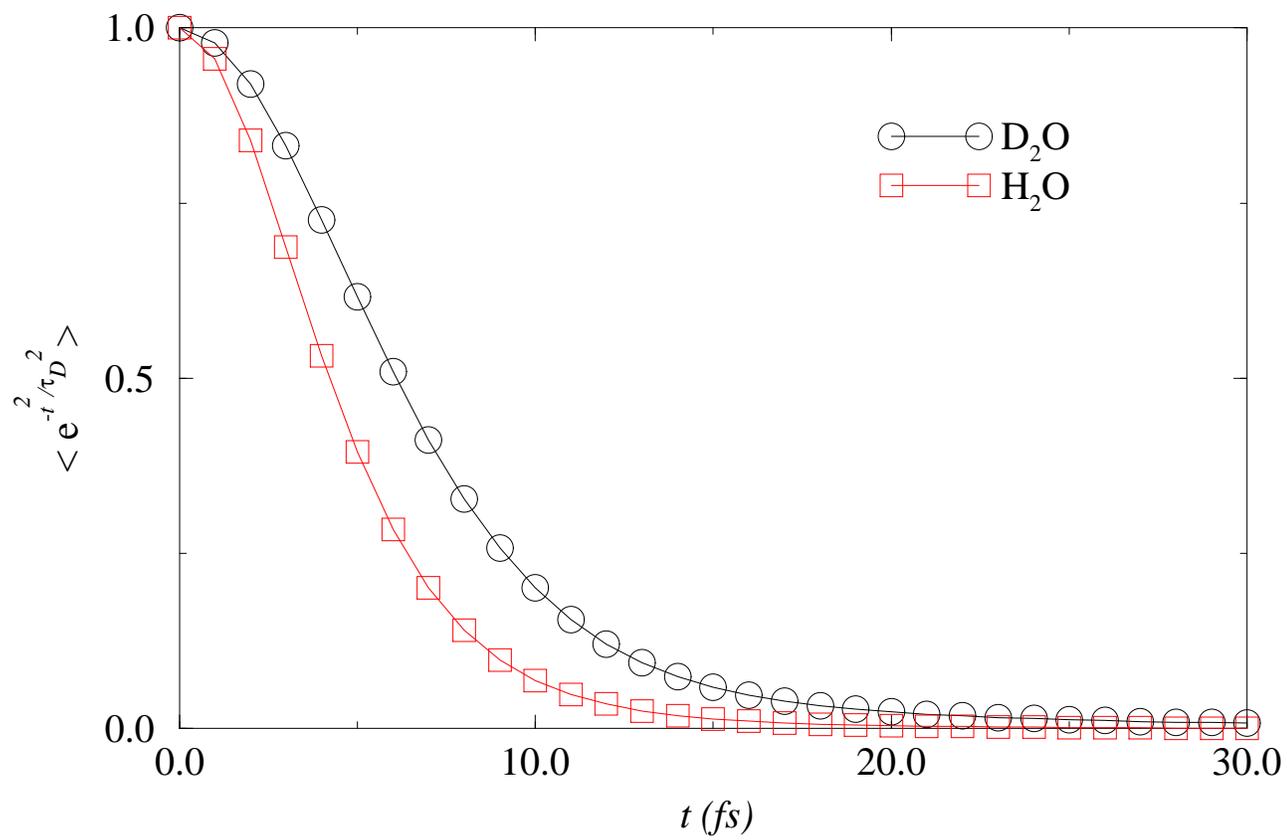
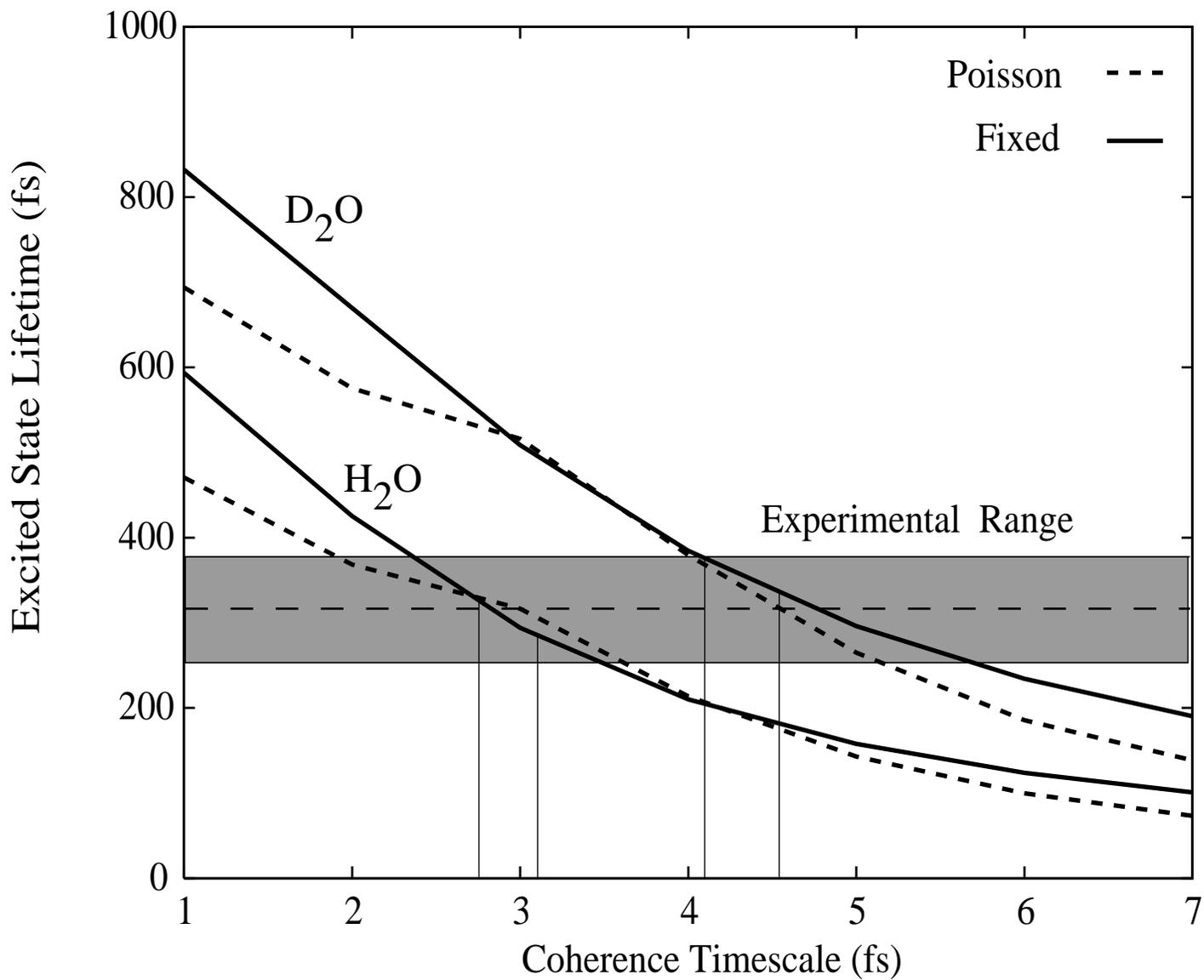
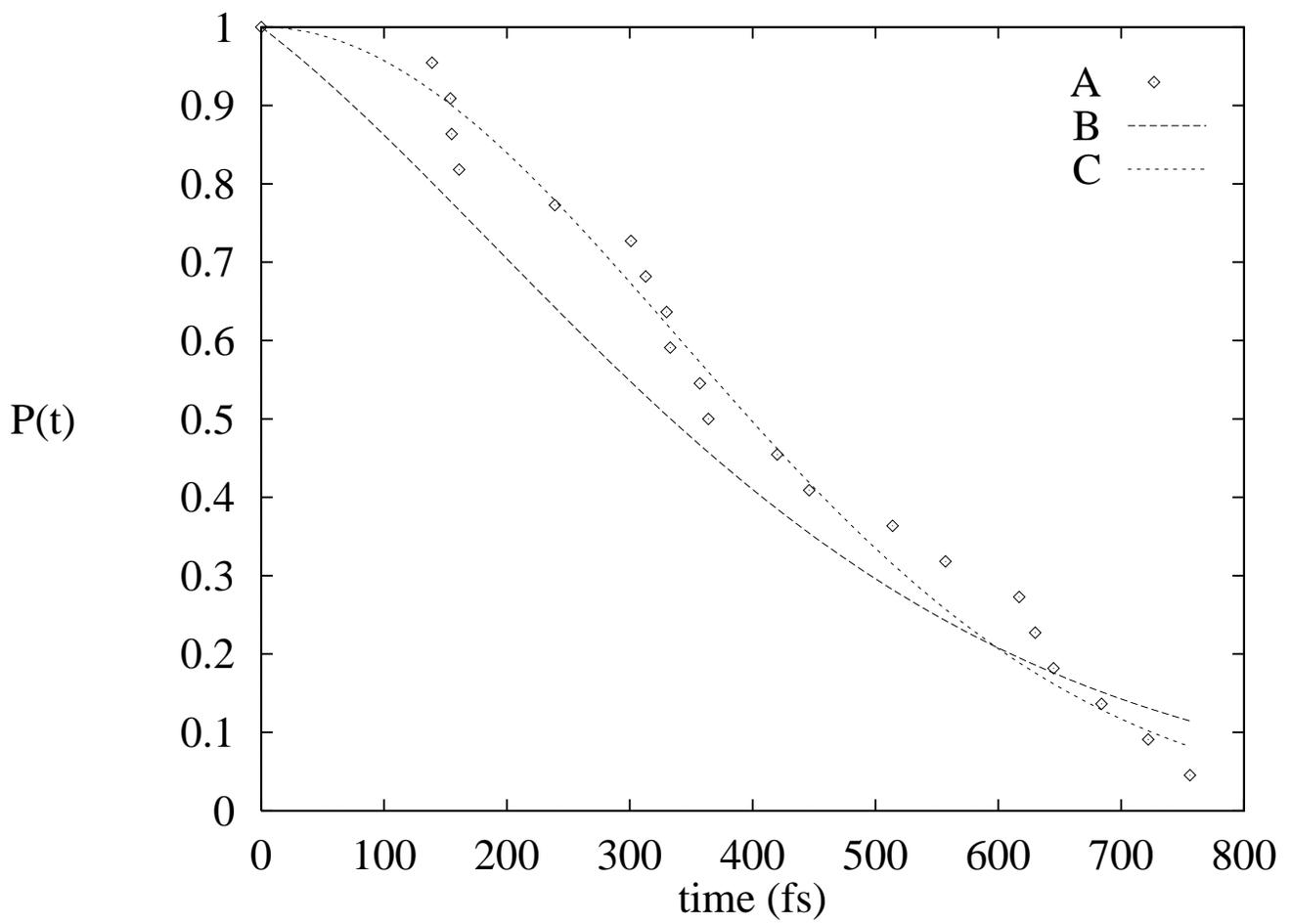
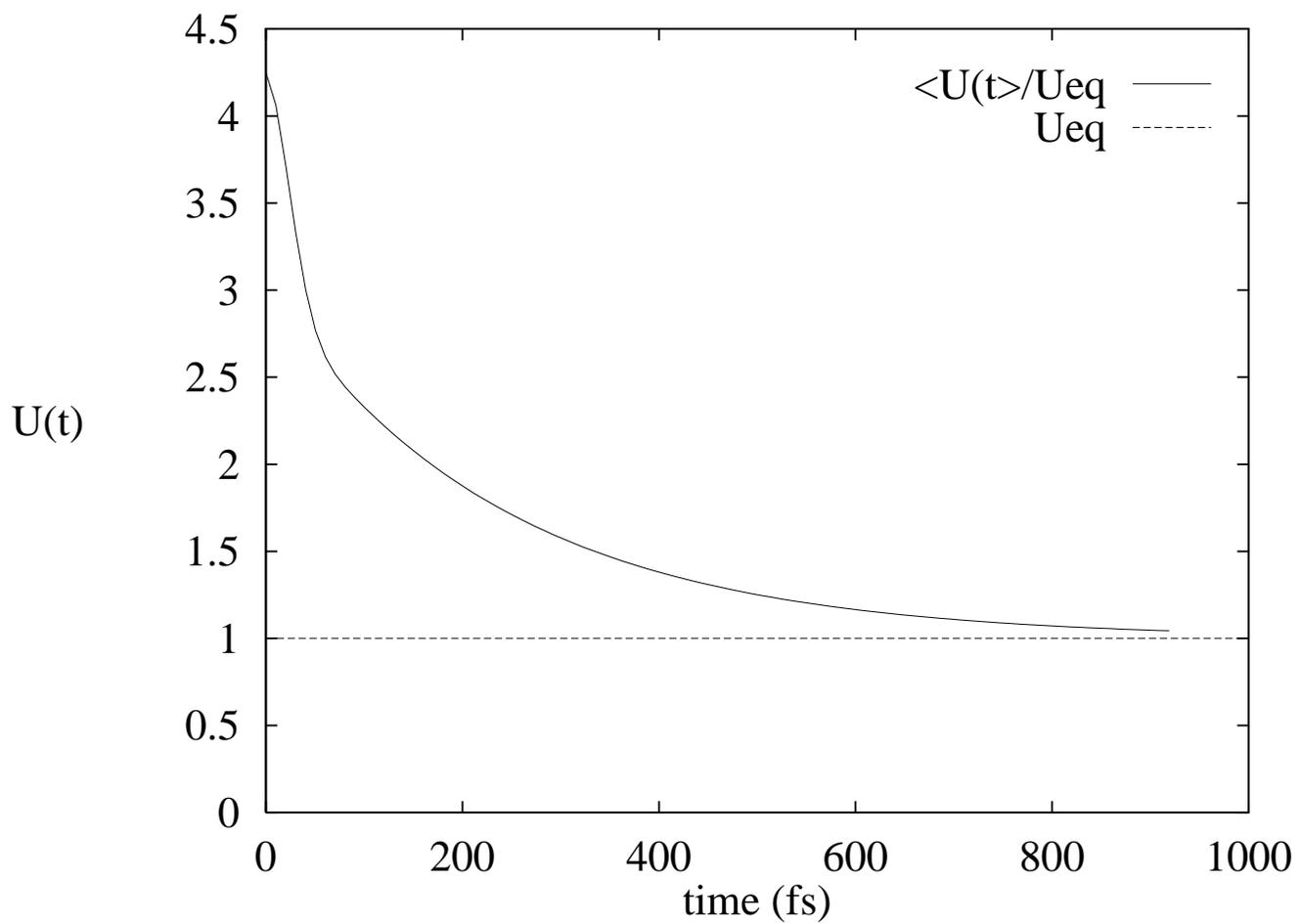
\end{document}